\documentclass[a4paper,twocolumn,preprintnumbers,floatfix,superscriptaddress,aps,prx,showpacs,notitlepage,longbibliography]{revtex4-2}

\usepackage[english]{babel}
\usepackage{ucs}
\usepackage[utf8x]{inputenc}
\usepackage{graphicx}
\usepackage{dcolumn}
\usepackage{bm}
\usepackage{bbm}
\usepackage{amsmath}
\usepackage{amssymb}
\usepackage{mathrsfs}
\usepackage{amsfonts}
\usepackage[table, dvipsnames]{xcolor}
\usepackage[colorlinks=true,citecolor=PineGreen,linkcolor=PineGreen,urlcolor=PineGreen]{hyperref}
\usepackage[normalem]{ulem}
\usepackage[ruled,vlined,linesnumbered]{algorithm2e}
\usepackage{algorithmic}
\usepackage[draft,inline,nomargin]{fixme}

\DeclareMathOperator*{\argmax}{arg\,max}

\newcommand{\ket}[1]{\left|#1\right\rangle}

\begin{document}

\title{Usefulness of Quantum Entanglement for Enhancing Precision in Frequency Estimation}
\author{Marco A. Rodr\'iguez-Garc\'ia}
\affiliation{Center for Quantum Information and Control, Department of Physics and Astronomy, University of New Mexico, Albuquerque, New Mexico 87131, USA}
\affiliation{Instituto de Investigaciones en Matemáticas Aplicadas y en Sistemas, Universidad Nacional Autónoma de México, Ciudad Universitaria, Ciudad de. México 04510, Mexico}
\author{Ruynet L. de Matos Filho}
\affiliation{Instituto de F\'isica, Universidade Federal do Rio de Janeiro, Rio de Janeiro, RJ 21941-972, Brazil}
\author{Pablo Barberis-Blostein}
\affiliation{Instituto de Investigaciones en Matemáticas Aplicadas y en Sistemas, Universidad Nacional Autónoma de México, Ciudad Universitaria, Ciudad de. México 04510, Mexico}

\begin{abstract}
  We investigate strategies for reaching the ultimate limit on the precision of
  frequency estimation when the number of probes used in each run of the
  experiment is fixed. That limit is set by the quantum Cram\'er-Rao bound
  (QCRB), which predicts that the use of maximally entangled probes enhances the
  estimation precision, when compared with the use of independent probes.
  However, the bound is only achievable if the statistical model used in the
  estimation remains identifiable throughout the procedure. This in turn sets
  different limits on the maximal sensing time used in each run of the
  estimation procedure, when entangled and independent probes are used. When
  those constraints are taken into account, one can show that, when the total
  number of probes and the total duration of the estimation process are counted
  as fixed resources, the use of entangled probes is, in fact, disadvantageous
  when compared with the use of independent probes. In order to counteract the
  limitations imposed on the sensing time by the requirement of identifiability
  of the statistical model, we propose a time-adaptive strategy, in which the
  sensing time is adequately increased at each step of the estimation process,
  calculate an attainable error bound for the strategy and discuss how to
  optimally choose its parameters in order to minimize that bound. We show that
  the proposed strategy leads to much better scaling of the estimation
  uncertainty with the total number of probes and the total sensing time than
  the traditional fixed-sensing-time strategy. We also show that, when the total
  number of probes and the total sensing time are counted as resources,
  independent probes and maximally entangled ones have now the same performance,
  in contrast to the non-adaptive strategy, where the use of independent is more
  advantageous than the use of maximally entangled ones.
\end{abstract}

\maketitle
\section{Introduction}

Frequency estimation is of fundamental interest as an essential ingredient for
experimental tests of our physical theories
\cite{PhysRevLett.113.210801,PhysRevD.91.015015, Barontini2022} and serves as a
cornerstone for numerous practical applications, including magnetometers
\cite{danilin2018quantum, dong2021high}, gravimeters \cite{wu2019gravity,
  stray2022quantum}, and atomic clocks
\cite{macieszczak2014bayesian,sanner2019optical, PhysRevX.9.041052}. To estimate
an unknown frequency that is encoded in a quantum system, it is necessary to
measure an observable of the system and use the resulting data as input for an
estimator function. The output of the estimator function represents an estimate
of the frequency. The theory of parameter estimation aims to find measurement
strategies and estimators that lead to small estimation errors. Quantum
Mechanics, on one side, sets limits on how small these erros can be and, on the
other side, establishes what are the resources necessary to reach those limits.
The ultimate quantum limit on the estimation error is given by the so-called
quantum Cramér-Rao bound (QCRB) \cite{Braunstein1994}, and the foremost
objective in quantum parameter estimation is the development of strategies that
can reach this bound.

In this article, we focus our attention on the spectroscopy of two-level systems
(qubits), which has an important example in the estimation of the frequency of
an atomic transition. The basic metrological scheme to estimating the frequency
$\omega$, using $N$ two-level probes, consists in first preparing the $N$ probes
in an adequate quantum state, letting them evolve for a time $t$ and then
measuring some observable on the final state. The measurement results are fed
into an estimator function, which produces an estimate of $\omega$.
Specifically, when the $N$ probes are prepared in an initial product state, the
QCRB is given by
\begin{equation}
  \label{eq:QCRB_frequency}
  \text{Var}_{\omega}(\widehat{\omega}(X_1,\ldots,X_N)) \geq \frac{1}{N t^2_\textrm{prod}}\, ,
\end{equation}
where $X_1, \ldots , X_N$ is a sample of the outcomes of $N$ independent
measurements, each on one of the $N$ probes, $\widehat{\omega}(X_1,...,X_N)$ is
an unbiased estimator, which satisfies $\text{E}_\omega\left[
  \widehat{\omega}(X_1,...,X_N) \right] = \omega$, $t_\textrm{prod}$ is the
evolution (sensing) time of the product state, and $\rm{Var}_{\omega}\left(
  \cdot \right)$ is the estimator variance \cite{Bollinger,Cirac1997,Cohen2020}.
The inverse of the right hand side in Eq.~\eqref{eq:QCRB_frequency} is equal to
the quantum Fisher information, which is a measure of how much information about
the parameter $\omega$ can be extracted from the sample $(X_1,\ldots, X_N)$. If
the results of $\nu$ repetitions of the experiment are taken into account, the
QCRB assumes the form
\begin{equation}
  \label{eq:QCRB_frequency2}
  \text{Var}_{\omega}(\widehat{\omega}(\{X_1^{(i)},\ldots,X_N^{(i)}\}_{i=1}^\nu)) \geq
  \frac{1}{\nu N t^2_{\textrm{prod}}}\, ,
\end{equation}
where ${X_1^{(i)}, \ldots ,X_N^{(i)}}$ represents a sample of size $N$, obtained
in the experiment $i$. It is well known that, in the asymptotic limit for the
number of measurements ($\nu \to \infty$), the maximum likelihood estimator
(MLE), derived from a sample of measurements that saturate the quantum
information bound (as described below), achieves the saturation of the
inequality Eq.~\eqref{eq:QCRB_frequency2}, thus producing the minimum possible
estimation error for this situation\cite{Toscano2017}. Note that the total time
required to satisfy the bound given by Eq.~\eqref{eq:QCRB_frequency2} is at
least $\nu t_{\textrm{prod}}$, while the total number of probes required is at
least $\nu N$. Looking at Eq.~\eqref{eq:QCRB_frequency2}, it is evident that the
error can be reduced by increasing either the number $N$ of probes in the
initial product state, the number of times $\nu$ the experiment is repeated, or
the evolution (sensing) time $t_{\textrm{prod}}$.

At first sight, since the variance decreases as the inverse of a quadratic
polynomial in time and of a linear polynomial in $\nu N$, one can conclude that
an efficient way to made the error as small as needed is by increasing
$t_{\textrm{prod}}$. Nevertheless, as noted in Ref. \cite{bonato2016optimized},
there is a limit on how much $t_{\textrm{prod}}$ can be increased because the
information on the frequency is generally codified into the relative phase
$\phi=(\omega - \omega_0) t_{\textrm{prod}}$, between the ground state
($\ket{g}$) and the excited state ($\ket{e}$) of the probe, where $\omega_0$ is
a known frequency (the ``clock'' frequency). Since the relative phase is
$2\pi$-periodic, the probabilities of the outcomes of measurements made on the
probe become $\pi$-periodic in $\phi$. That is, all phase values $\phi + n\pi$,
with $n \in \mathbb{Z}$, produce the same measurement outcomes. This implies
that it does not exist a MLE which can make a unique estimation (the likelihood
functions are not identifiable) and the Cramér-Rao bound can not be saturated
\cite{Robert,Lehmann1998}. A solution to this non-identifiability problem is to
restrict the value of $\phi$ to $-\pi/2\leq \phi \leq \pi/2$. In other words,
since some prior knowledge on the value of the frequency $\omega$ is typically
present, if one assumes that this value lies inside some interval around the
value of $\omega_0$, say $ \omega\in
(\omega_0-\Delta\Omega,\omega_0+\Delta\Omega)$, then the sensing time must be
restricted to $t_{\textrm{prod}} \leq \pi/(2\Delta\Omega)$. This restriction has
the side effect that one cannot choose arbitrary large sensing times to decrease
the estimation error.

It is well known that the use of quantum resources may lead to improvement of
error bound (\ref{eq:QCRB_frequency2}). Preparing the probes in a maximally
entangled state, a so-called GHZ state~\cite{greenberger1989going}, instead of
in a product state, gives rise to a smaller QCRB
\cite{Bollinger,escher2011quantum}. Specifically, when $N_{\textrm{GHZ}}$ probes
are prepared in a initial GHZ state, the quantum Cramér-Rao bound becomes
\begin{equation}
\label{eq:HLintroduction}
\text{Var}_{\omega}\left(\widehat{\omega}(\{X^{(i)}\left( N_{\textrm{GHZ}} \right)  \}_{i=1}^\nu) \right) \geq \frac{1}{\nu N^2_{\textrm{GHZ}} t^2_{\textrm{GHZ}}}\, ,
\end{equation}
where $t_{\textrm{GHZ}}$ is the sensing time of the GHZ state. It is noteworthy
that the variance bound Eq.~(\ref{eq:HLintroduction}), for GHZ states, scales as
$1/N^2_{\textrm{GHZ}}$, in contrast to the scaling as $1/N$ of
bound~(\ref{eq:QCRB_frequency2}). This scaling is known as the Heisenberg limit
\cite{Cirac1997} and is the best scaling possible in frequency
estimation~\cite{Bollinger}. When considering multiple probes and equal sensing
times, the bound given by Eq.\eqref{eq:HLintroduction} is smaller than the bound
in Eq.\eqref{eq:QCRB_frequency2}. Therefore, if both bounds can be saturated,
using GHZ states offers an advantage over using product states for achieving the
smallest possible estimation error. However, the problem of non-identfiability
in the likelihood functions plays an important role here. As it will be
discussed below, for an initial GHZ state, the information on the frequency
$\omega$ will be encoded in the relative phase
$\phi_{\textrm{GHZ}}=N_{\textrm{GHZ}}(\omega-\omega_0)t_{\textrm{GHZ}}$. This
phase evolves with the sensing time $t$ faster than the relative phase of an
initial product state: $\phi_{\textrm{GHZ}}=N_{\textrm{GHZ}}\cdot\phi$. If, as
in the case of an initial product state, one assumes that the value of the
frequency $\omega$ lies inside an interval around $\omega_0$, say $ \omega\in
(\omega_0-\Delta\Omega,\omega_0+\Delta\Omega)$, then , in order to maintain the
identifiability of the estimator, the sensing time $t_{\textrm{GHZ}}$ must be
restricted to $t_{\textrm{GHZ}} \leq \pi /(2N_{\textrm{GHZ}}\Delta\Omega)$. This
means that, for an inicial GHZ state, the maximal sensing time
$t_{\textrm{GHZ}}$ must be $N_{\textrm{GHZ}}$ times shorter than the
corresponding maximal sensing time $t_{\textrm{prod}}$, for an initial product
state. If these restrictions on the sensing times are taken into account in the
bounds given in Eqs.~(\ref{eq:QCRB_frequency2}) and (\ref{eq:HLintroduction}),
it becomes evident that the advantage of using maximally entangled states,
stemming from the quadratic scaling with the number of probes in the estimation
precision, can be canceled by the scaling with the inverse of the number of
probes of the corresponding maximal sensing time.

Indeed, as will be detailed below, when the estimation procedure consists of a
large number of preparation-sensing cycles, with a fixed sensing time, there is
no advantage in using maximally entangles states of the probes for improving the
precision of frequency estimation, when compared with the use of product states.
This result, which may seem counterintuitive, is consequence of the necessity of
limiting the sensing time in a way that allows one to get an identifiable
statistical model (set of parametrized probability distributions where different
values of the frequency must generate different probability distributions) and
then a consistent estimator (the estimates converge in probability to the
unknown parameter). These conditions are indispensable for saturating the
Cramér-Rao bound \cite{Fujiwara2011,RodriguezGarcia2021efficientqubitphase}.

The problem of non-identifiability in the statistical model also occurs in
interferometry (phase estimation), which is closely related to frequency
estimation, and has been discussed by several authors.
Ref.~\cite{pezzesmerzi2007}, in particular, presents a phase estimation
procedure with maximally entangles estates of the probes, where the variance of
the estimator may scale as $1/N^2_T$, with $N_T$ being the total number of
probes used in each run of the experiment. Their procedure, however, requires
that, in each run of the experiment, $p$ independent measurements be performed
on different sets of $N_p$ probes prepared in an initial maximally entangled
state, where $N_p=1,2,4,...,2^{p-1}$. In the limit of very large $p$, the
results of the $p$ measurements can be combined to construct a consistent phase
estimator, whose variance scales as $1/N^2_T$, where $N_T=2^{p-1}$ is the total
number of probes used in each run of the experiment. The adaptation of this
procedure to frequency estimation is straightforward and has been described in
Ref.~\cite{Oh_2014} and implemented in Ref.~\cite{bonato2016optimized}. For
phase measurements with the electromagnetic field, Ref.~\cite{Berry2009} solves
the problem of non-identfiability by proposing the use of an adaptive
measurement on multiple copies of NOON states distributed in multiple time
modes. In their scheme, one first perform $M$ measurements, one on each of $M$
NOON states with the same number $\nu=2^K$ of photons. Subsequently,
measurements are performed on $M$ NOON states with $\nu=2^{K-1}$, and this
sequence of measurements continues on NOON states with $\nu=2^k$ for $k=K,
K-1,\cdots,1$.

The above schemes successfully get rid of the phase non-identification problem, but
their practical implementation is a very challenging task, since a sequence of
maximally entangled states with different number of probes has to be prepared
and used in each run of the experiment. For this reason, we would like to
investigate in this article adaptive estimation strategies that use the same
initial state of the probes in each run of the experiment in order to solve the
non-identifiability problem in frequency estimation. When $\nu$ experiments are
done to estimate a parameter, adaptive methods consist of using the observed
data from previous measurements to choose the next quantum measurement, using a
suitable cost function. Under the right conditions, such methods surpass the
performance of the non-adaptive ones \cite{paninski2005asymptotic,
  Fujiwara2011}. In fact, adaptive estimation strategies have become a powerful
tool for overcoming limitations in diverse estimation problems
\cite{Fujiwara2011, Boxio, Huang2017, Berry2001,rodriguez2022determination,
  RodriguezGarcia2021efficientqubitphase}.

Some of the authors have recently demonstrated the benefits of adaptive
techniques to overcome the non-identifiability problem for phase estimation. For
instance, in the case of single-shot phase estimation in coherent states, an
adaptive estimation technique that leverages photon counting, displacement
operations, and feedback has been developed. It enables the avoidance of the
non-identifiability problem in likelihood functions and, by optimizing the
correct cost function, surpasses the performance of Gaussian measurements
\cite{rodriguez2022determination}. Additionally, an adaptive estimation
procedure based on confidence intervals has been proposed for the problem of
phase estimation in two-level systems. This approach overcomes the
non-identifiability limitation of likelihood functions produced by locally
optimal measurements and enables the saturation of the QCRB in the asymptotic
limit \cite{RodriguezGarcia2021efficientqubitphase}.

In this article, we present an adaptive strategy that solves the
non-identifiability problem for frequency estimation, in which the probes are
prepared in the same initial quantum state in each run of the experiment. This
strategy allows one to increase the measurement time in bounds
\eqref{eq:QCRB_frequency2} and \eqref{eq:HLintroduction}, decreasing the
frequency estimation error without non-identification problems. We calculate an
error bound for this strategy and discuss how to optimally choose its parameters
in order to minimize that bound. Finally, we discuss the possible advantage of
using maximally entangled states of the probes, when compared with the use of
the probes prepared in an initial product state.

\section{Identifiability limits: the usefulness of GHZ states}

In this section, we show in more detail how the need for an identifiable
statistical model cancels any possible advantage of using initial maximally
entangled states of the probes for frequency estimation, when the estimation
procedure consists of a large number of preparation-sensing cycles with a fixed
sensing time.

Any protocol for estimating the transition frequency of a two-level system can
be related to a standard Ramsey spectroscopy procedure. First, each one of $N$
two-level probes, of transition frequency $\omega$, is prepared in its ground
state $\ket{g}$ and a Ramsey pulse of frequency $\omega_ 0$ is applied to them.
Shape and duration of this pulse are such that each probe is put in a balanced
superposition of its ground state $\ket{g}$ and its excited state $\ket{e}$.
Next, the probes evolve freely for a time $t$ (sensing time) followed by a
second Ramsey pulse of same shape and duration of the first pulse. Finally, the
internal state of each one is measured and the probability of finding a probe in
the excited state $\ket{e}$ is $P=\cos^2[(\omega-\omega_0)t/2]$. This protocol
is repeated $\nu$ times and the measurement results are used to estimate the
frequency $\omega$.

In order to be possible to associate each value of $P$ to a single
value of $\omega$, the phase $\phi=(\omega-\omega_0)t/2$ must lie
inside an interval of width smaller than half the width of a fringe of
$\cos^2[(\omega-\omega_0)t/2]$. Since one typically has some prior
knowledge on the value of $\omega$, if it is assumed that this value
lies inside some interval around $\omega_0$, for example
$ \omega\in (\omega_0-\Delta\Omega,\omega_0+\Delta\Omega)$, that
condition can be met only if the sensing time $t$ is restricted to
$t\leq \pi/(2\Delta\Omega)$. In this case, using the maximum allowed
sensing time $t_\textrm{prod} = \pi/(2\Delta\Omega)$, the minimum
error possible in the estimation of $\omega$ is obtained from the
quantum Cramér-Rao inequality~\eqref{eq:QCRB_frequency2}:
\begin{equation}
  \label{eq:QCRB_frequency_max_ramsey}
  {\rm Var}_{\omega}(\widehat{\omega}(\{X_1^{(i)}, \ldots ,X_N^{(i)}\}_{i=1}^\nu)) \geq \frac{4\Delta\Omega^2}{\pi^2\nu N
  }\, ,
\end{equation}
with the use of a total number $\nu N$ of probes. It is natural to
consider the total time $T$ of the estimation procedure as a resource.
Assuming that the duration of the Ramsey pulses is much shorter than
the sensing time $t$ of each run of the experiment, one can set
$T=\nu t$. Inserting this relation in
Eq.~\eqref{eq:QCRB_frequency_max_ramsey}, the minimum error in the
estimation of $\omega$, when using the maximum sensing time
$t_{\rm prob}=\pi/ (2\Delta\Omega)$, is determined by
\begin{equation}
  \label{eq:QCRB_totaltime_max}
  {\rm Var}_{\omega}(\widehat{\omega}(\{X_1^{(i)}, \ldots ,X_N^{(i)}\}_{i=1}^\nu))\geq
  \frac{2\Delta\Omega}{\pi T N}\, ,
\end{equation}
showing that the variance of $\omega$ scales as $1/(T N)$. To reach this limit,
a total number of at least $\nu N$ probes is needed.

It is well known that, if the probes are put in an initial maximally entangled
state, the QCRB for the estimation precision of the frequency $\omega$ becomes
smaller, leading, in principle, to a smaller error in the estimation of $\omega$
\cite{Bollinger}. For this purpose, each one of $N_{\textrm GHZ}$ probes is
initialized in its ground state $\ket{g}$. A Raman pulse is then applied to one
of the probes followed by a set of controlled-NOT operations on the other
probes. This prepares the probes in a maximally entangled state
\begin{equation}
\ket{\psi}=\frac{1}{\sqrt{2}}\left(\ket{gg\cdots g}+\ket{ee\cdots e}\right).
\end{equation}
Next, the probes evolve freely for a time $t$ followed by a second Ramsey pulse
on the same probe that received the fist pulse and a set of controlled-NOT
operations on the other probes. Finally, the internal state of the probe that
received the Ramsey pulses  is measured and the probability of finding it in the excited state
$\ket{e}$ is $P=\cos^2[N_{\textrm GHZ}(\omega-\omega_0)t/2]$. This protocol is
repeated $\nu$ times and the measurement results are used to estimate the
frequency $\omega$ \cite{Cirac1997}.

Notice that the probability $P$ oscillates $N_{\textrm GHZ}$ times faster than
the probability corresponding to initial product states, leading to the improved
quantum Cram\'er- Rao bound~\eqref{eq:HLintroduction}. Comparing
bounds~\eqref{eq:QCRB_frequency2} and \eqref{eq:HLintroduction}, one can see
that, for the same sensing time $t$ and equal numbers $\nu$ of runs, the use of
$N$ probes in an initial maximally entangled state may lead to an improvement in
the precision of the estimation of $\omega$ by a factor $\sqrt{N}$. This clearly
shows the advantage of the use of entangled states for improving the bound on
the estimation of the frequency $\omega$. However, in order to maintain the
phase $\phi_{\textrm GHZ}= N_{\textrm GHZ}(\omega-\omega_0)t/2$ inside an interval of width smaller than a half width of a
fringe of $P$, the sensing time $t$ must be limited. If one assumes, like in the
case of an initial product state of the probes, that the frequency $\omega$ lies
inside the interval $(\omega_0-\Delta\Omega,\omega_0+\Delta\Omega)$, the
sensing time must be restricted to $t\le \pi/(2N_{\textrm GHZ}\Delta\Omega)$.
Using the maximum allowed sensing time $t_{\textrm GHZ}=\pi/(2N_{\textrm
  GHZ}\Delta\Omega)$ in bound~\eqref{eq:HLintroduction}, results in
\begin{equation}
  {\rm Var}_{\omega}\left(\widehat{\omega}(\{X^{(i)}\left( N_{\textrm{GHZ}} \right)  \}_{i=1}^\nu) \right)\geq
  \frac{4\Delta\Omega^2}{\pi^2\nu}\label{eq:noaqsescale}\, .
\end{equation}
This bound does not even depend on the number $N_{\textrm GHZ}$ of probes in the
entangled initial state. This is a consequence of the fact that, to
guarantee a consistent estimator, it is not possible to increase the number
$N_\textrm{GHZ}$ of probes in the entangled state without decreasing the sensing
time $t_{\textrm GHZ}$ by the same proportion.

Comparison of bounds~\eqref{eq:noaqsescale} and
\eqref{eq:QCRB_frequency_max_ramsey} shows that the use of initial maximally
entangled states is disadvantageous when compared with the use of initial product
states of the probes, if the only resource taken into account is the total
number $\nu N=\nu N_{\textrm GHZ}$ of probes. Indeed, the maximal reachable
precision in the estimation of $\omega$ is worse by a factor $\sqrt{N}$ .

Since the sensing time $t$, when using probes in a product state, can be much
larger than the sensing time when using maximally entangled states, it is
essential to count the total time $T=\nu t$ of the experiment as a resource. In
this case, bound~\eqref{eq:noaqsescale} becomes
\begin{equation}
  \label{eq:QCRB_totaltime_ghz_max}
  {\rm Var}_{\omega}\left(\widehat{\omega}(\{X^{(i)}\left( N_{\textrm{GHZ}} \right)  \}_{i=1}^\nu) \right)\geq \frac{2\Delta\Omega}{\pi T N_\textrm{GHZ}}\, .
\end{equation}
Notice that the total number of probes necessary to reach this limit is at least
$\nu N_\textrm{GHZ}$. Bound~\eqref{eq:QCRB_totaltime_ghz_max} should be compared
with bound~\eqref{eq:QCRB_totaltime_max} for initial product states. For equal
total duration $T$ of the estimation procedures and equal number $N=N_{\textrm
  GHZ}$ of probes used in each run, they set the same upper limit to the
precision of the estimation of $\omega$. However, in each run of the estimation
process, the maximum sensing time $t_{\textrm GHZ}$ allowed with the use of
$N_{\textrm GHZ} $ entangled probes is shorter than the maximum sensing time
$t_{\textrm prob}$, optimal for the use of $N=N_{\textrm GHZ}$ independent
probes, by a factor $N_{\textrm GHZ}$. Since $T$ is the same in both cases, this
implies that the number $\nu$ of runs with the use of entangled probes has to be
$N_{\textrm GHZ}$ times larger than the number $\nu$ of runs with the use of
independent probes. This, on the other side, implies that the total number $\nu
N$ of probes used in the former case is larger than the total number of probes
used in the latter case. In the limit of large number of measurements, $\nu\gg 1$,
the two bounds are saturated and set the best precision effectively reachable in
both cases. Consequently, when the total number $\nu N$ of probes and the total
duration $T$ of the estimation process are counted as fixed resources, the use
of initial GHZ states of the probe is disadvantageous when compared with the use
of independent probes. In order to reach the same precision within a fixed total
time $T$, the strategy that uses GHZ states needs $N_{\textrm GHZ}$ more probes
than the strategy that uses product states.

\begin{figure}[h!]
  \centering
  \includegraphics[width=3.4in]{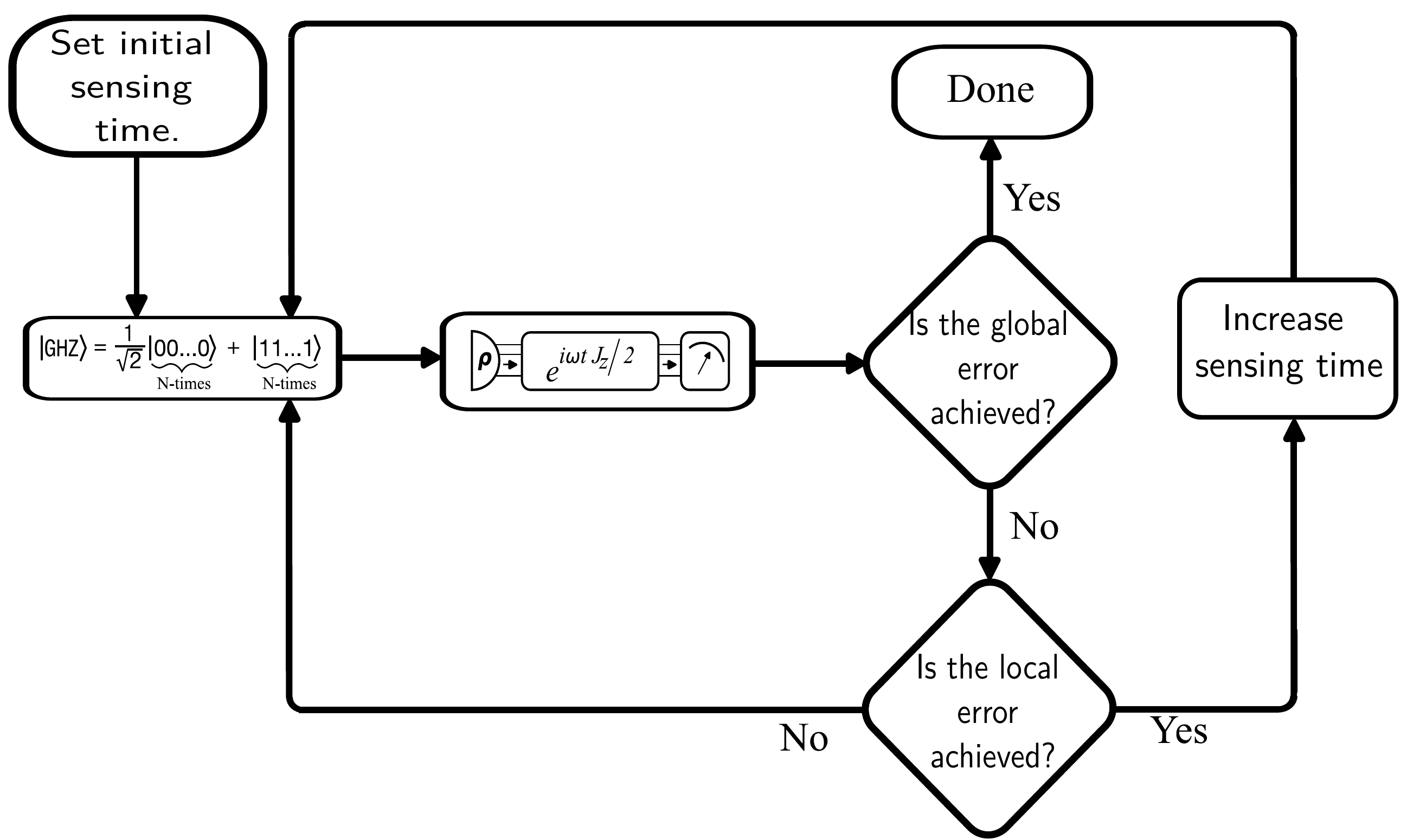}
  \caption{\textbf{Adaptive-time frequency estimation}. The procedure begins with
    a measurement of a phase encoded in a quantum state. If the measurement
    error is smaller than the desired threshold (local error) for the current
    step, the measurement time is increased, the threshold is recalculated, and
    the process is repeated. In this case, the quantum Fisher information
    available for the subsequent measurement increases. Conversely, if the
    measurement error is equal to or exceeds the current threshold, the
    measurement is repeated without adjusting the measurement time. This
    procedure is repeated until the desired overall error (global error) is
    achieved.}
  \label{fig:Scheme}
\end{figure}

The above discussion makes it clear that, if the initial state of the
probes and the sensing time are fixed, the need for an identifiable
statistical model cancels any advantage of the use of entangled states
for estimation of the frequency $\omega$. In order to address the
limitation on the sensing time, we shall introduce an adaptive
estimation strategy that takes advantage of the fact that the maximum
sensing time increases as $1/\Delta\Omega$ when the frequency interval
shrinks. The key idea of this approach is to start with a small number
of identical preparation-sensing-measurement cycles to produce an
initial estimate of $\omega$. This first estimation allows one to
shorten the interval $\Delta\Omega$ and, consequently, to increase the
sensing time for the next set of measurements. This process is
repeated until the desired estimation error is achieved (see
Fig.~\ref{fig:Scheme}). We present the details of this strategy in the
next sections.

\section{Estimation with Confidence Intervals}

\label{sec:ECI}

In this section, we review frequency optimal measurements and the conditions that
lead to consistent estimators to propose and simulate a frequency estimation
strategy that minimizes the error by increasing the measurement time. We call the
proposed method Adaptive-Time Frequency Estimation (ATFE).

\subsection{Optimal measurements and asymptotic consistent estimators}

The goal is to estimate an unknown parameter $\omega$, codified into a
two-level system as a phase $\phi = (\omega-\omega_0) t$, where $t$
and $\omega_0$ are control parameters. Without loss of generality, the
codification process can be represented by an unitary operation on a
fiducial state of the two-level system
\begin{equation}
  \label{eq:rho_f}
  \rho(\omega, t) = \hat U_\omega(t) \rho \,\hat U^\dagger_\omega(t),
\end{equation}
with
$\hat U_\omega(t)=e^{-i (\omega-\omega_0)t \frac{\vec{\sigma}}{2}
  \cdot \vec{n} }$. Here, $\vec{\sigma}$ is the vector of Pauli
matrices, $\vec{n}$ a unit vector and
$\rho = \frac{1}{2}\left(I + \vec{\sigma} \cdot \vec{a} \right)$ a
fiducial state of the two-level system, characterised by the Bloch
vector $\vec{a}$. The codification process leads to a new state, which
is represented by the Bloch vector resulting from the rotation of
$\vec{a}$, by the angle $(\omega-\omega_0)t$, around an axis parallel
to $\vec{n}$ .

To estimate $\omega$ we use an estimator $\widehat{\omega}: \mathcal{X} \to
\Omega$, which is a function whose domain is defined in the set of measurement
outcomes $\mathcal{X}$. Since the frequency is codified in the relative phase
between the two levels of the system, $\rho(\omega, t)$ is periodic in $\omega$,
and as a result, the estimations would also be periodic in $\omega$. To ensure
uniqueness in the estimation, the range of the estimator's values should be
defined inside an interval smaller than that period.

The general description of measurements on a quantum system is given by Positive
Operator Valued Measures (POVMs) \cite{Holevo2011,helstrom1969quantum}. Given an
outcome space $\mathcal{X}$ and the Borel $\sigma$-algebra $\mathcal{B}\left(
  \mathcal{X} \right)$ that represents the events that can be observed in an
experiment, a POVM with outcome space $\mathcal{X}$ is a $\sigma$-additive map
(countable additivity for a sequence of pairwise disjoint events) $P:
\mathcal{B}\left( \mathcal{X} \right) \to \mathcal{B}\left( \mathcal{H} \right)$
from the Borel $\sigma$-algebra to the space $\mathcal{B}\left( \mathcal{H}
\right)$ of bounded operators on $\mathcal{H}$
\cite{Holevo2011,beneduci2011relationships}. In the case
of a finite outcome space $\mathcal{X}$, the set of positive operators $P(k) $
on $\mathcal{H}$ with the property that $\sum_{\mathcal{X}} P(k) = I$, where $I$
is the identity operator on $\mathcal{H}$, determines a positive operator valued
measure (POVM). Hence, given a measurement on a state $\rho(\omega)$ of a
quantum system, with outcomes $x \in \mathcal{X} \subset \mathbb{R}$ and
described by a POVM $P = \left\{ P(x) \mid x \in \mathcal{X} \right\}$, the
conditional probability distribution for $x$ is given by Born's rule
\begin{equation}
  \label{eq:born_rule}
  p(x \mid \omega) = \textrm{Tr}\left[P(x) \rho(\omega) \right].
\end{equation}
Thus, given a sample $X$ of results from the application of the
    POVM $P$, the expected value of any estimator $\widehat{\omega}\left( X
    \right)$ based on this sample is defined as
\begin{equation}
  \label{eq:ev_est}
  \text{E}_{\omega}\left[\widehat{\omega}\left( X \right)  \right] = \sum_{x \in \mathcal{X}} p(x \mid \omega) \widehat{\omega}(x).
\end{equation}

It can be the case that the estimator is a multivalued function. For
instance, as noted above, the frequency estimator that uses
measurements from a two-level system is periodic. This periodicity
introduces a problem when calculating the estimation error. For
example, given $t$, if $\hat\omega_P$ is the period in frequency of
the quantum state \eqref{eq:rho_f}, two possible estimations for the
same experimental data are $\omega+\epsilon$ and
$\omega+\epsilon+m\,\hat\omega_P$, with $m$ an integer and
$\epsilon\ll 1$; one of these estimations has an error $\epsilon$,
whereas the other has an error $\epsilon+m\,\hat\omega_P$.

To correctly calculate that variance, it is necessary to modify the cost
function to account for the fact that estimations that differ by a multiple of
the period correspond to the same value of the parameter to be estimated. In such
scenarios, an appropriate measure is given by the Holevo variance of
$\widehat{\omega}(X)$ \cite{Holevo2011,Berry2001}:
\begin{equation}
  \label{eq:var_est}
  \textrm{Var}_{\omega}\left[ \widehat{\omega}(X) \right] =  (\frac{\hat\omega_P}{2\pi})^2\Big(\Big\lvert
    \sum_{x \in \mathcal{X}} p(x \mid \omega) e^{2\pi \mathrm{i} (\widehat\omega(x))/\hat\omega_P}
  \Big\rvert^{-2}-1\Big)\, .
\end{equation}
This variance reduces to the usual definition when the error is small.

When a POVM $P=\left\{ P(x) \mid x \in \mathcal{X} \right\}$ is
performed on the state $\rho(\omega)$, the minimum attainable
estimation error of any unbiased periodic estimator
$\widehat{\omega}\left( X \right)$ of the frequency $\omega$ is
bounded by the Cramér-Rao bound (CRB)
\begin{equation}
  \label{eq:CRB}
  \textrm{Var}_{\omega}\left[ \widehat{\omega}\left( X \right) \right] \geq \frac{1}{F(\omega; P)}.
\end{equation}
Here, $F(\omega; P)$ represents the Fisher information that a sample $X$ of results from
the POVM $P$ carries about the parameter $\omega$.  It is computed from the
probability distribution of $X$ \cite{Holevo2011,Lehmann1998}:
\begin{equation}
  \label{eq:FI}
  F(\omega; P) = \textrm{E}_{\omega} \left[ \left(
      \frac{\partial}{\partial \omega} \log \left( p(x \mid \omega)
      \right)    \right)^2  \right].
\end{equation}
Particularly, for a sample $X_1, \ldots X_{\nu}$ of size $\nu \geq 1$, obtained
from $\nu$ identically and independently applied measurements $P$, the CRB for
any unbiased estimator $\widehat{\omega}_\nu := \widehat{\omega}\left(
  X_1,\ldots,X_\nu \right)$ based on this sample is given by
\begin{equation}
  \label{eq:CRB_N}
  \textrm{Var}_{\omega}\left[ \widehat{\omega}_\nu \right] \geq \frac{1}{\nu
    F(\omega; P)}\, .
\end{equation}
This inequality follows from the additivity of the Fisher information
\cite{degroot}. The estimators that saturate the CRB are called efficient, and
when this condition is met in the asymptotic limit ($\nu \to \infty$), they are
called asymptotically efficient. A well-known result in statistics is that,
under a set of regularity conditions, the MLE is an asymptotically efficient
unbiased estimator \cite{CaseBerg:01,Fujiwara2011}.

The ultimate limit of precision in quantum mechanics is achieved by optimizing
the quantity $F(\omega; P)$ over all POVMs. This optimization process yields the
quantum Fisher information (QFI) about the unknown parameter $\omega \in \Omega
\subseteq \mathbb{R}$. The QFI is a function of $\omega$ that is independent of
any specific POVM and is defined as \cite{Holevo2011,helstrom1969quantum}:
\begin{equation}
  \label{eq:QFI_def}
  F_{\textrm{Q}}(\omega) = \text{Tr}\left[ \rho(\omega) \lambda(\omega)^2 \right] \,,
\end{equation}
where $\lambda(\omega)$ is the symmetric logarithmic derivative (SLD), which is
implicitly defined by the equation
\begin{equation}
  \label{eq:SLD}
  \frac{d \rho(\omega) }{d \omega}  = \frac{1}{2}\left( \lambda(\omega)\rho(\omega) + \rho(\omega)\lambda(\omega) \right)\,.
\end{equation}
The QFI provides a generalization of the Cramér-Rao bound to the
quantum domain, the so-called quantum Cramér-Rao bound (QCRB)
\begin{equation}
  \label{eq:QCRB_2}
  \textrm{Var}_{\omega}\left[ \widehat{\omega}(X) \right] \geq \frac{1}{F_{\rm{Q}}(\omega)} \,.
\end{equation}
Additionally, the Fisher information of a POVM $P$ satisfies
\cite{Braunstein1994}
\begin{equation}
  \label{eq:qib}
  F(\omega; P) \leq F_{\textrm{Q}}(\omega)\, .
\end{equation}
This inequality is known as the quantum information bound (QIB). The
POVMs that saturates Eq.~(\ref{eq:qib}) are called optimal and are the
most sensitive measurements for the estimation of the parameter. A
sufficient condition for achieving the QIB is given by the POVM
$P_{\rm{L}}(\omega)$, whose elements are the projectors onto the
eigenspaces of the SLD operator \cite{Braunstein1994, Toscano2017}. By
construction, this POVM depends on $\omega$, which is the parameter to
be estimated (it is locally optimal). For this reason, if one
considers $\left\{P_{\rm{L}}(g) \right\}_{g\in\Omega}$ as a one
parameter family of POVMs, one can only guarantee that $P_{\rm{L}}(g)$
achieves the QIB if $g=\omega$, turning the method useless because the
value of $\omega$ is unknown. However, in some systems, one can find
initial conditions where any POVM $P_{\rm{L}}(g)$ saturates
Eq.~(\ref{eq:qib}) for every $g \in \Omega$ and independently of
$\omega$~\cite{Toscano2017}.

In the context of frequency estimation, the quantum Fisher information (QFI) of
the state Eq.~(\ref{eq:rho_f}) is independent of $\omega$ and is given as:
\begin{equation}
  \label{eq:QFI_qub_general}
 F_{\rm{Q}} = t^2\left[ 1 -\left( \vec{a} \cdot \vec{n} \right)^2 \right] \, .
\end{equation}
When the Bloch vector associated with the initial state of the probe
is perpendicular to the rotation axis of the frequency codification
process, $\vec{n} \cdot \vec{a} = 0$, the quantum Fisher information
$F_{\rm{Q}}$ reaches its maximum value $t^2$. Moreover, under this
initial conditions, any POVM $P_{\rm{L}}(g)$ saturates
Eq.~(\ref{eq:qib}) for any $\omega \in \Omega$ \cite{Chapeau2016,
  RodriguezGarcia2021efficientqubitphase}. We will assume this initial
condition from now on. Explicitly, the elements for any
$P_{\rm{L}}(g) = \left\{ P(x;g) \mid x \in \left\{ 0,1 \right\}
\right\}$ are
\begin{equation}
  \label{eq:P_{rm{L}}}
  \begin{split}
    &P(0;g) = \frac{1}{2}\left( I + \vec{n}  \times \vec{a}(g)  \cdot \vec{\sigma} \right), \\
    &P(1;g) = \frac{1}{2}\left( I - \vec{n}  \times \vec{a}(g)  \cdot \vec{\sigma} \right),
 \end{split}
\end{equation}
where $a(g) = \cos((g-\omega_0) \,t) \vec{a} + \sin((g-\omega_0)\, t) \vec{n} \times \vec{a}$, $g \in
\Omega$. Thus, according to the Born's rule
\begin{equation}
  \label{eq:p_pl}
  p(x \mid \omega; t) = \begin{cases} \frac{1}{2}\left[ 1+ \sin((\omega - g)\,t  ) \right] \, \text{if } x=0 \\
                          \frac{1}{2}\left[ 1- \sin((\omega - g) \,t ) \right] \, \text{if } x=1
  \end{cases}
\end{equation}
and from Eq.~(\ref{eq:FI}),
\begin{equation}
  \label{eq:FI_pl}
  F(\omega; P_{\textrm{L}}(g)) = t^2 = F_{\textrm{Q}}(\omega), \, \, \, \forall g, \omega \in \Omega.
\end{equation}
To exemplify how a physical measurement is codified in the POVM
\eqref{eq:P_{rm{L}}} we use a specific example considering that a
two-level system is algebraically equivalent to the electron spin. Let
the initial condition of the probe be a state with spin in the $x$
direction, described by the Bloch vector $a=(1,0,0)$, and the
codification process be a rotation around the $y$ direction,
$\vec{n}=(0,1,0)$. If $g=\omega_0$, from Eq.~ \eqref{eq:P_{rm{L}}} it
is easy to see that the elements of the POVM $P_{\rm{L}}(g)$ describe
a measurement of the component of the spin in the $z$ direction.

As noted above, the MLE obtained by a sequence of $\nu$ independent outcomes of
any $P_{\rm{L}}(g)$ can saturate the CRB under a set of regularity condition
\cite{degroot}. For frequency estimation, one of the regularity conditions that
fails to be satisfied is that the likelihood function has to be identifiable.
For example, using the POVM \eqref{eq:P_{rm{L}}} the likelihood function is
\begin{equation}
  \label{eq:lik}
  L(\omega) = p(0 \mid \omega; t)^{m}\left( 1 - p(0 \mid \omega; t) \right)^{\nu-m},
\end{equation}
where $m$ is the number of $0$s in the measured data. Inserting
Eq.~\eqref{eq:p_pl} in the above expression, it is straightforward to see that
this likelihood function has multiple global maxima, separated of each other by
$\pi$, rendering it asymptotically inconsistent and biased. A way to achieve
saturation of the CRB is to render the likelihood function \eqref{eq:lik}
identifiable by restricting its image to a sufficiently small interval where
there is only one global maximum.

\begin{figure}[htb]
  \centering
  \includegraphics[width=3.3in,height=2.9in]{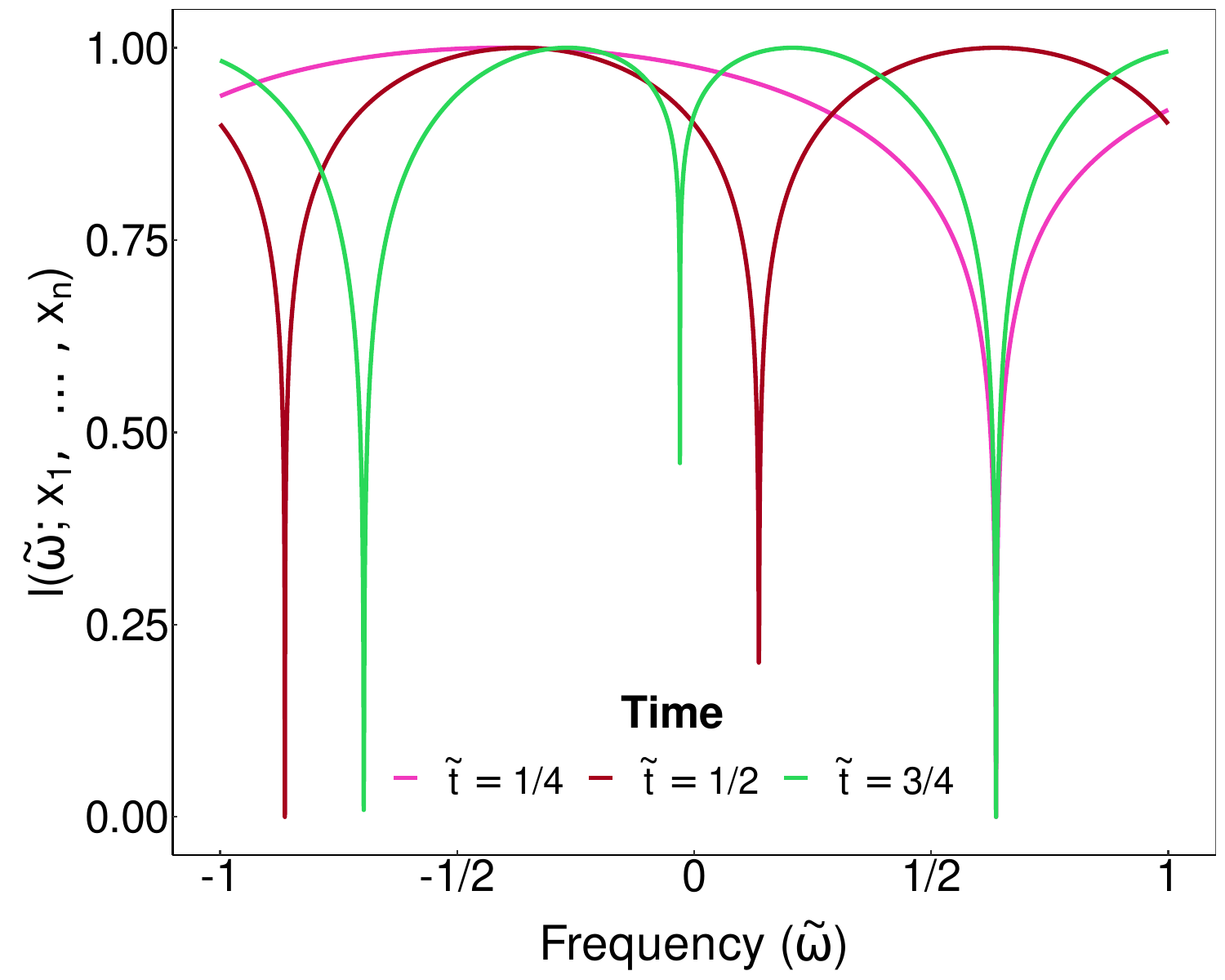}
  \caption{\textbf{Normalized log-likelihood functions produced by
      $P_{\rm{L}}(\tilde{g})$}. When $\tilde{t}=1/4$ the likelihood
    function is identifiable. For $\tilde{t}=1/2$ and $\tilde{t}=3/4$
    there are $2$ and $3$ maxima, respectively, making the likelihood
    functions not identifiable. For all cases, the likelihood
    functions were plotted using $64$ data from
    $P_{\rm{L}}(\tilde{g})$ and assuming that $\tilde{\omega} = 2$.}
  \label{fig:Qubit}
\end{figure}

Summarizing, the POVM~\eqref{eq:P_{rm{L}}} saturates the QIB,
Eq.~\eqref{eq:qib}, but two problems arise if one wants to use it to
minimize the estimation error: i) it is locally optimal and ii) for
large measurement times its associated likelihood, Eq.~(\ref{eq:lik}),
becomes non-identifiable. The first problem can be avoided using the
optimal initial conditions or adaptive estimation schemes
\cite{Fujiwara2011}; the second one can be avoided by requiring an
unique estimation in $\omega\in\Omega$, which can be accomplished if
$t \leq \pi/(2\Delta\Omega)$, where $2\Delta\Omega$ is the
length of the interval defined by $\Omega$. Fig.~\ref{fig:Qubit} shows
an example of how the number of maxima increases by one if $t$
increases by $\pi/(2\Delta\Omega)$, making it impossible to
find a consistent frequency MLE for $t > \pi/(2\Delta\Omega)$.

Non-identifiability problems arises in qubit phase estimation, where
the locally optimal POVM produces a likelihood with two maxima. In
\cite{RodriguezGarcia2021efficientqubitphase} this non-identifiability
was solved using an adaptive estimation technique based on the
construction of confidence intervals from a prior sample of the
canonical phase measurement \cite{Holevo1979,Martin2019} and a
posterior adaptive sequence implementation of the locally optimal
POVM. One of the objectives of this paper is to find an adaptive
estimation strategy that allows one to increase the measurement time
in the bound \eqref{eq:QCRB_frequency2}, decreasing the frequency
estimation error in two-level systems.

To introduce the proposed strategy, we first define confidence
intervals as a set of plausible values that are likely to contain the
true parameter value, with the confidence level representing the
proportion of such intervals containing the parameter's value, in the
limit of an infinite number of repeated experiments. Using the MLE
$\widehat{\omega}_{\textrm{MLE}}(X)$ for a sample $X$, an estimator
for a confidence interval with confidence level
$0 \leq C_l=1-\alpha \leq 1$ can be computed as follows:
\begin{widetext}
\begin{equation}
  \label{eq:conf_intervals}
    \widehat{\textrm{CI}}(\widehat{\omega}_{\textrm{MLE}}(X)) =
    \left( \widehat{\omega}_{\textrm{MLE}}(X)  -
     z_{\alpha/2} \cdot F(\widehat{\omega}_{\textrm{MLE}}(X) )^{-\frac{1}{2}}, \widehat{\omega}_{\textrm{MLE}}(X)
    + z_{\alpha/2} \cdot F(\widehat{\omega}_{\textrm{MLE}}(X))^{-\frac{1}{2}} \right)\ ,
 \end{equation}
\end{widetext}
where $F(\omega)$ is the Fisher information of the sample, $z_{\alpha}$ is the
$\alpha$-ith quantile of the standard normal distribution ($P(Z \geq z_{\alpha})
=\alpha$) \cite{CaseBerg:01}. Hence, every outcome of
$\widehat{\textrm{CI}}(\widehat{\omega}_{\textrm{MLE}}(X))$ is a possible
confidence interval.

Eq.~\eqref{eq:conf_intervals} assumes that the distribution of the MLE is
asymptotically normal, with its mean at $\omega$ and variance equal to the
inverse of the Fisher information. This requires the use of estimation
strategies with consistent estimators. Thus, the identifiability of the
parameters becomes crucial, since it is both a sufficient and necessary
condition for obtaining an asymptotically consistent MLE
\cite{CaseBerg:01,Robert}.

\subsection{Minimizing the estimation error using adaptive
  measurements and confidence intervals: Adaptive-Time Frequency Estimation (ATFE)}

A quick explanation of the method is as follows: we start by finding a set
$\Omega$ such that $\omega \in \Omega$ and setting a measurement time $t_1$ such
that the likelihood function given in Eq.~(\ref{eq:lik}) has a single stationary
point as a global maximum within that region. Subsequently, we perform an
optimal POVM to obtain the data $x_1$ and an estimation
$\widehat{\omega}_{\textrm{MLE}}(x_1)$. By using this estimate, a confidence
interval $CI_1 = \widehat{\textrm{CI}}(\widehat{\omega}_{\textrm{MLE}})(x_1)
\subset \Omega$ is obtained, according to Eq.~(\ref{eq:conf_intervals}).
Assuming that $\omega \in CI_1$, we set a new measurement time $t_2 > t_1$ such
that the statistical model restricted to $CI_1$ remains identifiable. After
repeating the experiment, a new frequency estimation inside a smaller confidence
interval $CI_2$ is obtained. By iteratively performing this procedure, we
increase the measurement time at each adaptive step until the desired error is
achieved.

To describe the method in detail, one first assumes that
$\Omega=[\omega_0-\Delta\Omega,\omega_0+\Delta\Omega]$ and defines an
adimensional time, $\tilde{t}=t\Delta\Omega/2\pi $, with the goal to estimate
the adimensional frequency $\tilde{\omega}=(\omega-\omega_0)/\Delta\Omega$. With
this notation $\tilde{\omega}\in\tilde{\Omega}=[-1,1]$. For $N$ probes initially
prepared in a product state, the choice of the maximal measurement time
$\tilde{t}=1/4$ guarantees that
\begin{equation}
  \label{eq:scaling}
  -\pi/2 \leq (\omega-\omega_0) t \leq \pi/2\, ,
\end{equation}
whereas for $N$ probes initially prepared in a GHZ state, the above condition is
satisfied when $\tilde{t}=1/(4N)$. The relation between the variance of a
estimator for the adimensional frequency $\tilde{\omega}$ and an estimator for
the actual frequency $\omega$ is
\begin{equation}
  \label{eq:relation_fisher}
 \textrm{Var}\left(\widehat{ \omega} \right)= \Delta\Omega^2\cdot\textrm{Var}\left( \widehat{\tilde{\omega}} \right)\,.
\end{equation}
For simplicity, throughout the rest of the paper, $\widehat{\omega}$ will be an
estimator for $\tilde{\omega}$.

For a single probe, one starts by selecting a measurement time
$\tilde{t}_1=1/4$. This yields a likelihood function with a single global
maximum in $\tilde{\Omega}$, when the parameter $g$ in the POVM
Eq.~\eqref{eq:P_{rm{L}}} is close to $\tilde\omega$. Then, the first estimation
strategy, termed $M_1$ and described in detail in the following, is employed.
This strategy consists of making measurements on $\nu_1$ independent probes,
each of them producing a Fisher information
$F_{Q_1}=(2\pi\tilde{t}_1)^2=\pi^2/4$, until the total fisher information
$F_1(\tilde{\omega})$, obtained after the $ \nu_1$ measurements, satisfies
$z_{\alpha/2} \cdot F(\widehat{\omega}_{\textrm{MLE}}(x_1)
)^{-\frac{1}{2}}\approx 1/2$, where $x_1$ is the data set containing the
measurements results. The confidence interval for the outcome $x_1$ is then (see
\eqref{eq:conf_intervals}) $CI_1=\left(\widehat{\omega}_{\textrm{MLE}}(x_1)-1/2,
  \widehat{\omega}_{\textrm{MLE}}(x_1)+1/2 \right)$, where
$\widehat{\omega}_{\textrm{MLE}}(x_1)$ is the estimation using the strategy
$M_1$. The subsequent step is to extend the measurement time to
$\tilde{t}_2=2\tilde{t}_1$, thereby increasing the Fisher information that can
be gained in each measurement. Notice that increasing the measurement time
introduces a second maximum in $\tilde{\Omega}$ that is displaced by $1$ with
respect to the first maximum. This second maximum can turn the likelihood
function non identifiable inside the set $\tilde\Omega$. However, if the first
strategy was successful, this second maximum lies outside $CI_1$. Therefore,
restricting the next estimation strategy $M_2$ to the set $CI_1$ guarantees that
the likelihood function remains identifiable. The probability that the parameter
is not in $CI_1$ is $1-P(\theta \in CI_1)=\alpha$. In general, the $i$th
strategy $M_i$ consists of $\nu_i$ measurements that generates a sample $X_i$,
and $\widehat{\omega}_{\textrm{MLE}}(X_1,X_2,\ldots,X_i)$. By choosing the
Fisher information of the $M_i$ strategy such that $z_{\alpha/2} \cdot
F(\widehat{\omega}_{\textrm{MLE}}(X_1,X_2,\ldots,X_i) )^{-\frac{1}{2}}\approx
1/(i+1)$ we obtain an estimator for the confidence interval
\begin{equation}
  \begin{split}
&\widehat{CI}_i = \\ &\left(
  \widehat{\omega}_{\textrm{MLE}}(X_1,X_2,\ldots,X_i) -
  1/(i+1), \right. \\ &\widehat{\omega}_{\textrm{MLE}}(X_1,X_2,\ldots,X_i) +
    1/(i+1) \left. \right)\, ,
 \end{split}
\end{equation}
which allows one to increase the measurement time to
$\tilde{t}_i=i \tilde{t}_1$ for the next measurement strategy,
increasing the Fisher information gained in each measurement to
$F_{Q_i}(\tilde{\omega})=(\pi i)^2/4$ and diminishing the estimation error. In
order to find the bound for the estimation error of the whole
procedure we have to specify the estimation strategies $M_i$.

To obtain a performance close to the QCRB, it is necessary to obtain an
asymptotic efficient estimator (asymptotically unbiased, with variance equal to
the inverse of Fisher information). To this end, we use the Adaptive Quantum
State Estimation method (AQSE) \cite{Fujiwara2011}, in the region $CI_i$, with
the family of POVMs $\left\{ P_{\textrm{L}}(\tilde{g}) \right\}_{\tilde{g} \in
  \tilde{\Omega}}$, as the $i$-th optimally local estimation strategy $M_i$
\cite{Fuji2, Fujiwara2011}, with $\tilde{g}=(g-\omega_0)/\Delta\Omega$. Notice
that each strategy $M_i$ comprises $\nu_i$ measurement steps. The AQSE begins
with an arbitrary initial guess $\tilde{g}_0$. At this point, the locally
optimal measurement $P_{\rm{L}}(\tilde{g}_0)$ is applied. Assuming that the
outcome $x_1$ is observed, one obtains the likelihood function
$L_1(\tilde{\omega} ; x_1; \tilde{g}_{0}) = p(x_1\mid \tilde{\omega};
\tilde{g}_{0})$ and applies the MLE to produce an estimate
$\tilde{g}_1=\widehat{\omega}_1(x_1) = \argmax_{\tilde{\omega} \in \Omega}
L_1(\tilde{\omega} ; x_1; \tilde{g}_{0})$, which will be used as the new guess
for the POVM $P_{\rm{L}}$. Thereby, for the step $n \geq 2$, one applies the
POVM $P_{\textrm{L}}(\tilde{g}_{n-1} =
\widehat{\omega}_{n-1}(x_1,...,x_{n-1}))$, where
$\widehat{\omega}_{n-1}(x_1,...,x_{n-1}))$, is the estimation from the previous
stage, which used the outcomes $x_1,...,x_{n-1}$. If the outcome $x_n$, is
observed, the likelihood function at $x_n$ for step $n$ is
\begin{equation}
  \label{eq:like_AQSE}
  L_{n}(\tilde{\omega} ; x_1,...,x_{n}; \tilde{g}_{n-1}) = \prod_{i=1}^{n} p(x_i \mid
  \tilde{\omega}; \tilde{g}_{i-1})\, ,
\end{equation}
from which one gets the nth guess $\tilde{g}_n = \widehat{\omega}_{n}\left(
  x_1,...,x_n \right)$ by applying the MLE:
  \begin{displaymath}
    \widehat{\omega}_{n}(x_1,...,x_{n}) = \argmax_{\tilde{\omega} \in \tilde{\Omega}} L_{n}(\tilde{\omega} ; x_1,...,x_{n};
    \tilde{g}_{n-1})\, .
\end{displaymath}

Since $L_{n}$ is close to zero when $n \gg 1$, it is more convenient to use the
natural logarithm of the likelihood function, which has the same maxima than the
likelihood function. We denote the logarithm of the likelihood function by
$l(\tilde{\omega})$. An equivalent definition for the MLE of $\tilde{\omega}$,
given a sample $(X_1,...,X_n)$, is given by
\begin{equation}
  \widehat{\omega}_n(X_1,...X_n) = \argmax_{\tilde{\omega} \in \tilde{\Omega}} l(\tilde{\omega}; X_1,...,X_n).
\end{equation}
Since, for the optimal initial condition ($\vec{n}\cdot\vec{a}=0$), the Fisher
information $F_{Q_i}$ of each measurement step of the $i$th strategy does not
depend on the parameter $\tilde{g}$, the CRB for
$\widehat{\omega}_{n}(X_1,...X_n)$, produced from the AQSE method in $n$
adaptive steps, is
\begin{equation}
  \label{eq:AQSE_CRB}
  \text{Var}_{\omega}\left[ \widehat{\omega}_{n}(X_1,...,X_n)  \right] \geq \frac{1}{n F_{Q_i}} = \frac{1}{n(2\pi\tilde{t}_i)^2},
\end{equation}
for each strategy $M_i$.

It could be argued that AQSE is not necessary because the Fisher
information is the same regardless of the value of $\tilde{g}$.
However, AQSE is needed because it ensures a asymptotically normal
estimator \cite{Fujiwara2011,RodriguezGarcia2021efficientqubitphase},
which is a necessary condition to justify the use of Eq.
(\ref{eq:conf_intervals}), which define our confidence intervals.
Adopting AQSE as the $M_i$ estimation strategy implies that the
protocol of estimation with confidence intervals consists of two
adaptive processes: the primary adaptive process, with total Fisher
information $F_i$, which increases the measurement time after a
certain confidence interval is attained, and the secondary adaptive
process (with a fixed measurement time $\tilde{t}_i$ and Fisher
information $F_{Q_i}=(2\pi\tilde{t}_i)^2$ for each measurement), which
involves the implementation of AQSE as the $M_i$ estimation strategy,
wherein the POVM is altered after each measurement until the required
confidence interval is achieved.

We determine now the estimation error of ATFE. For a given confidence level $C_l
\in \left[ 0,1 \right]$, in each adaptive step, we have two contributions to the
estimation error: i) when the confidence interval
$\widehat{\textrm{CI}}=(\widehat\omega-E,\widehat\omega+E)$, where $E$ is the
marginal error and $\widehat\omega$ is the estimation, contains the value of the
parameter and ii) when it does not. When the confidence interval contains the
value of the parameter, the MLE from AQSE converges in distribution to a normal
distribution with mean at $\tilde{\omega}$ and variance equal to the inverse of
the Fisher information \cite{CaseBerg:01,Fujiwara2011}. On the other hand, if
the confidence interval does not contain the parameter, AQSE produces an error
larger than $E$ \cite{RodriguezGarcia2021efficientqubitphase}.

Thereby, given that $(C_l)^i$ and $(1-C_l)\sum_{j=1}^i (C_l)^{j-1}$ are the probability
that the parameter is inside and outside the confidence interval at
primary adaptive step $i$, respectively, the estimation error for a total
number $\nu=\sum_i^S \nu_i$ of measurements is
\begin{equation}
  \label{eq:var_method}
  \begin{split}
  &\text{Var}_{\omega}\left[ \widehat{\omega}_\textrm{MLE} \right] \geq
    \frac{(C_l)^S}{\sum_{i=1}^S  \nu_i F_{Q_i}}\\
    &+ \left( 1 -C_l \right)
    \left( E_1^2+C_l E^2_2+\cdots+(C_l)^{S-1} E^2_{S}\right),
  \end{split}
\end{equation}
where $S$ is the total number of primary time-adaptive steps, and $\nu_i$, the
number of measurements using time $\tilde{t}_i$ in the strategy $M_i$, is set by
the requirement that the resulting confidence interval be $CI_i$. When $\nu\gg
1$, the second contribution dominates in Eq.~(\ref{eq:var_method}), since
performing additional measurements does not reduces the error if the parameter
is outside of the confidence interval. For this protocol to be useful, the
confidence levels should be chosen in such a way that the second term of
Eq.~(\ref{eq:var_method}) is negligible. The marginal error is a function of the
Fisher information and the quantile of the standard normal distribution. To
estimate the minimum number $\nu_i^\textrm{min}$ of measurements at step $i$,
required to produce a confidence interval of length less than the marginal error
$E_i$, we use Eq.~\eqref{eq:conf_intervals}. From that equation it follows that
$E^2_i \geq z^2_{\alpha/2}/\sum_{j=1}^i \nu_j F_{Q_j}$, where $F_{Q_j}$ is the
Fisher information of each AQSE measurement of the estimation strategy $M_j$.
Since $F_{Q_j}=\pi^2j^2/4$ and $E_i=1/(i+1)$, this expression leads to the
condition $\sum_{j=1}^i\nu_j j^2\geq \frac{4}{\pi^2}
z^2_{\alpha/2}\left(i+1\right)^2$, which, for $i\geq 2$, can be rewritten as
$\nu_i^\textrm{min} i^2\geq \frac{4}{\pi^2}
z^2_{\alpha/2}\left(i+1\right)^2-\sum_{j=1}^{i-1} \nu_j^\textrm{min} j^2$. If
one substitutes $\sum_{j=1}^{i-1} \nu_j^\textrm{min} j^2$, on the right-hand
side of the former expression, by $\frac{4}{\pi^2} z^2_{\alpha/2}i^2$, one
finally gets
\begin{equation}
  \label{eq:mar_err}
  \nu_i^\textrm{min} \geq
  \frac{4}{\pi^2} z^2_{\alpha/2}\frac{2i+1}{i^2}.
\end{equation}
\begin{figure}[htb]
  \centering
  \includegraphics[scale = 0.33]{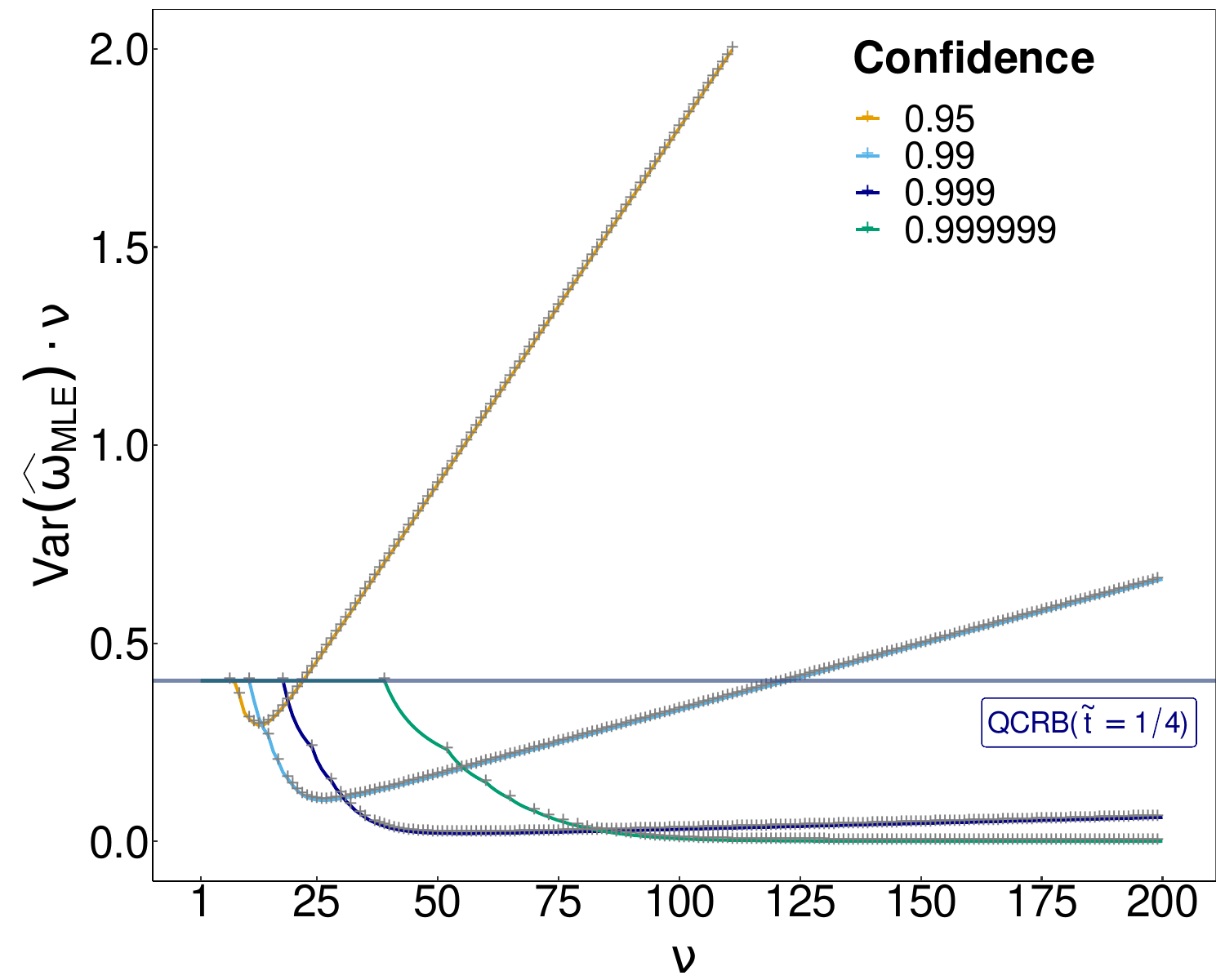}
  \caption{\textbf{Lower bound in the Holevo variance for the MLE produced by
      ATFE}. Eq.~(\ref{eq:var_method}) is plotted for the confidence levels $C_l
    = 0.95$ (yellow line), $C_l = 0.99$ (blue line), and $C_l = 0.999$ (green
    line). The cross markers in the curves indicate the adaptive step when the
    time is increased in the method. } \label{fig:Err_Qubit}
\end{figure}

The right-hand side of Eq.~\eqref{eq:mar_err} diminishes as a function of $i$
and will eventually become smaller than $1$ as the number of primary
time-adaptive steps increases. If we call $S_1$ the number of steps for which
this happens, then, from Eq.~\eqref{eq:mar_err}, it is easy to see that
$S_1\approx \frac{8}{\pi^2} z^2_{\alpha/2}$. As examples of typical values of
$S_1$, for a confidence level $C_l=0.99$, the value of $S_1$ will be $S_1\approx
5$, whereas for a confidence level $C_l=0.999999$, $S_1$ will increase to
$S_1\approx 19$. If one additionally assumes that the total number of
measurements at step $S_1$ is such that the MLE is already a normal estimator
that saturates the Crámer-Rao bound, then, when $i> S_1$, the choice of
$\nu_i=1$ will produce an error smaller than the marginal error goal for that
step. This means that the measurement time can be increased by $\tilde{t}_1$ at
each measurement, after step $S_1$. The behavior of the lower bound
Eq.~\eqref{eq:var_method} multiplied by the total number $\nu$ of measurements
is shown in Fig.~\ref{fig:Err_Qubit} for increasing confidence levels of the
confidence intervals. For comparison, the solid blue line shows the QCRB for a
fixed measurement time $\tilde{t}_i=1/4$, which corresponds to the minimum
possible error if no adaptive changes in the measurement times are allowed ( cf.
Eq.~\eqref{eq:QCRB_frequency2}). In this case $ \text{Var}_{\omega}\left[
  \widehat{\omega}_\textrm{MLE} \right]=4/(\pi^2\nu)$. From that figure it can
be seen that, as long as the second term of Eq.~\eqref{eq:var_method} is
negligible, the ATFE performs much better than the non-adaptive method. That
term, which is almost constant and limits the minimum error reachable in the
ATFE, depends only on the confidence level $C_l$ and on the confidence interval
widths $E_i$. Its value decreases when the confidence level $C_l$ increases;
this, in turn, increases the values of $\nu$ for which its contribution becomes
important. In section~\ref{sc:numsim}, we simulate the measurement and show that
the protocol reaches this bound.

\begin{figure}[h!]
  \centering \includegraphics[scale = 0.33]{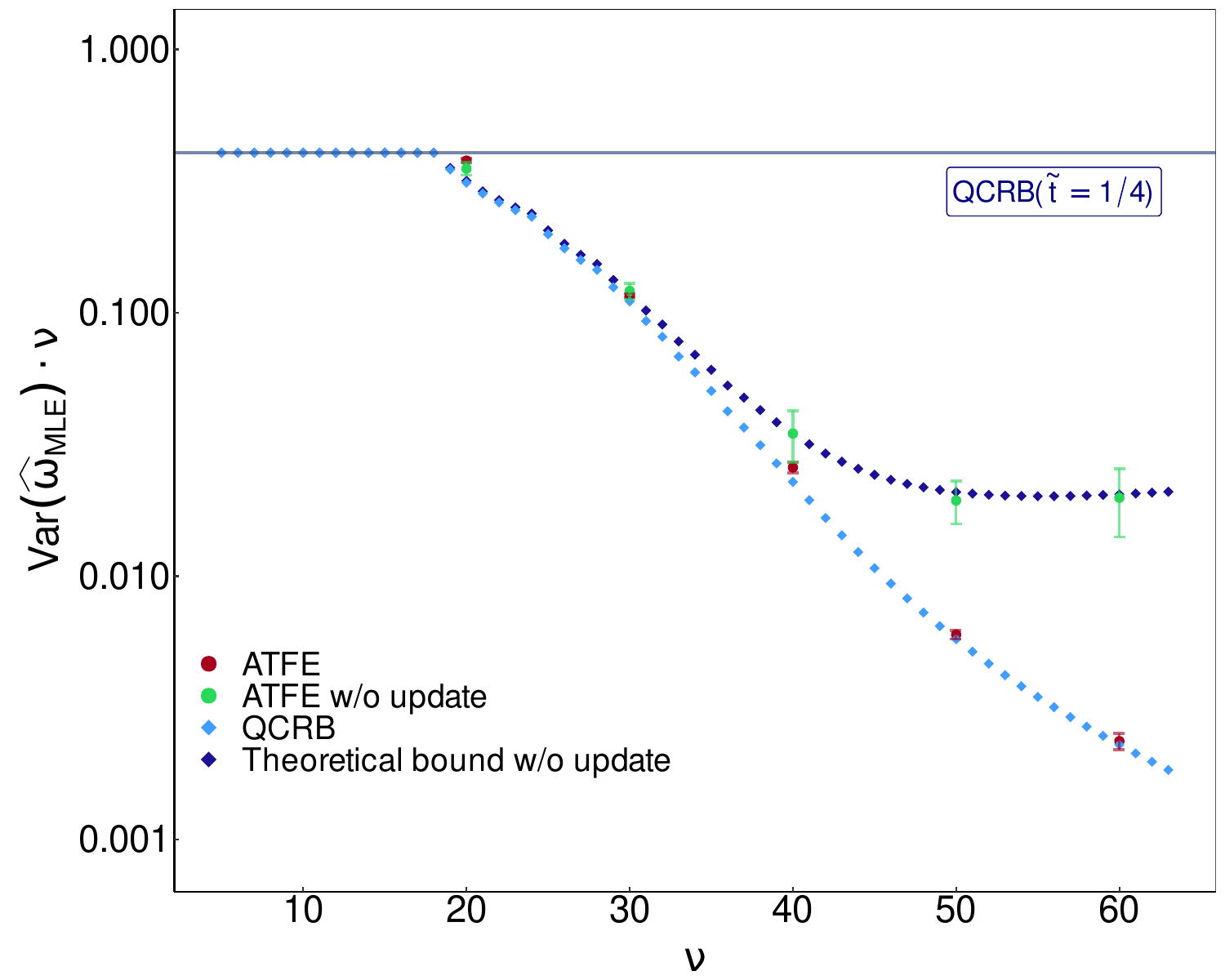}
  \caption{\textbf{ATFE Holevo variance vs the number of measurements for one
      qubit and a confidence level of $0.999$.} The lower bound error of the
    AQSE method for $\tilde{t}=1/4$ (gray line), for the ATFE without center
    update (green points) and for the ideal case of $C_l=1$ (light blue rhombus)
    are shown as reference. The dark blue rhombus is the error of the simulation
    of the ATFE strategy with no center update of the confidence interval at
    each step, it can be seen how the method can be used to diminish the
    estimation error for $\nu<40$. The red points is the error in the estimation
    of the simulation of the ATFE strategy with center update of the confidence
    intervals at each step, it can be seen how the method gives an estimation
    error close to the ideal bound for $\nu$ up to $60$. The expected Holevo
    variance at each point for both ATFE strategies was obtained from the
    average of 5 Monte Carlo simulations of the experiment, each with $10^3$
    samples, except for points $40$, $50$, and $60$ for ATFE w/o update
    strategy, where $10^4$ samples were used.}
  \label{fig:Numerical_Qubit}
\end{figure}

Now we focus on obtaining analytical approximations to
Eq.~\eqref{eq:var_method} to better understand how the error diminish
as a function of the resources. First, notice that the total number of
measurements at step $S$, $\nu=\sum_{i=1}^S \nu_i$, can be
approximated by $\nu=\sum_{i=1}^S \nu_i^\textrm{min}$. Using
Eq.~\eqref{eq:mar_err} for $i\le S_1$ and $\nu_i^\textrm{min}=1$ for
$i>S_1$, one obtains
\begin{equation}
\label{eq:nutot}
\nu=S_1\sum_{i=1}^{S_1}\frac{(i+1/2)}{i^2}+S-S_1\approx S_1\ln(S_1)+S,
\end{equation}
for $S>S_1$. In the same way, the value of
$F_S=\sum_{i=1}^S \nu_i F_{Q_i}$ can be approximated by
$F_S=\sum_{i=1}^S \nu_i^\textrm{min} F_{Q_i}$. If one considers the
case where each measurement is done on a product state with $N$
qubits, then $F_{Q_i}=N \pi^2 i^2/4$. Notice that, in this case, the
value of $ \nu_i^\textrm{min}$ is given by the value in
Eq.~\eqref{eq:mar_err} divided by $N$. The same happens to $S_1$.
Using Eq.~\eqref{eq:mar_err}, the value of $F_S$ is given by
\small{
\begin{eqnarray}
 F_S&=&N\frac{\pi^2}{4}\left[\frac{S_1}{2N}\sum_{i=1}^{S_1/N}(2 i+1)+\sum_{i=S_1/N+1}^S i^2\right]\nonumber\\
     &=&N\frac{\pi^2}{24}\left[\frac{S_1}{N}\!\left(\frac{S_1^2}{N^2}\!-\!3\frac{S_1}{N}\!-\!1\right)\!\!+\!\!S(S\!+\!1)(2S\!+\!1)\right],
\label{eq:Fs}
\end{eqnarray}}
\normalsize
when $S>S_1$. Using Eq.~\eqref{eq:nutot} and assuming that $S\gg 1$,
the above expression can be rewritten as
\small
\begin{equation}
  \label{eq:FsFin}
F_S\approx N\frac{\pi^2}{24}\left[\frac{S_1}{N}\!\left(\frac{S_1^2}{N^2}\!-\!3\frac{S_1}{N}\!-\!1\right)\!\!+\!\!2\left(\nu-\frac{S_1}{N}\ln(\frac{S_1}{N})\right)^3\right],
\end{equation}
\normalsize
where the value of $S_1$ is given by the value that follows from
Eq~\eqref{eq:mar_err}, which corresponds to the case of a single
probe. Inserting this expression into Eq.~\eqref{eq:var_method}, one
arrives finally at
\begin{widetext}
 \begin{equation}
   \label{eq:var_method_ideal}
  \text{Var}_{\omega}\left[ \widehat{\omega}_\textrm{MLE} \right] \geq
    \frac{24(C_l)^S}{ N\pi^2\left[\frac{S_1}{N}\!\left(\frac{S_1^2}{N^2}\!-\!3\frac{S_1}{N}\!-\!1\right)\!\!+\!\!2\left(\nu-\frac{S_1}{N}\ln(\frac{S_1}{N})\right)^3\right]}+\left( 1 -C_l \right)
    \left( \sum_{i=1}^{S}  C_l^{i-1} E_i^2\right)\, ,
 \end{equation}
\end{widetext}
for $S>S_1$ and $\nu\gg 1$. Assuming that the above inequality can be
saturated, the error in the estimation of $\tilde\omega$ decreases as
$1/\nu^3$ under the ATFE, as long as the second term is negligible,
which, compared with the non-time-adaptive strategy (cf.
Eq.~\eqref{eq:QCRB_frequency_max_ramsey}), is a huge advantage. This
behavior is shown in Fig.~\ref{fig:Numerical_Qubit}. As already noted,
the second term does not depend on $\nu$ and dominates when
$\nu\to \infty$, limiting the minimum error reachable in the ATFE. A
rough upper bound on that term can be stablished as
\begin{widetext}
\begin{equation}
\label{eq:boundsecterm}
\left( 1 -C_l \right)
     \sum_{i=1}^{S} ( C_l)^{i-1} E_i^2= \left( 1 -C_l \right) \sum_{i=1}^{S} \frac{(C_l)^{i-1}}{(i+1)^2}\leq
\left( 1 -C_l \right)\sum_{i=1}^{\infty} \frac{1}{(i+1)^2}\approx 0.64(\left(1-C_l\right).
\end{equation}
\end{widetext}
In section~\ref{sc:numsim} it will be numerically shown that the inequality in
Eq.~\eqref {eq:var_method_ideal} can be saturated.

We consider now the case where each measurement is done in a GHZ state
with $N_\textrm{GHZ}$ qubits. In this situation, as discussed before,
$\tilde t_1=1/(4N_\textrm{GHZ})$ and
$F_{Q_i}=N_\textrm{GHZ}^2\left(2\pi \tilde t_i\right)^2$. Using
$\tilde t_i=i \tilde t_1$ results in $F_{Q_i}=\pi^2 i^2/4$. This is
equal to the Fisher information obtained with measurements done on a
single probe. Consequently, the lower bound on the error in the
estimation of $\tilde\omega$ with GHZ states of $N_\textrm{GHZ}$
probes is given by substituting $N=1$ in
Eq.~\eqref{eq:var_method_ideal}, which is an improvement over
Eq.~(\ref{eq:noaqsescale}). On the other side, this implies that, if
only the total number of probes is counted as resource, product states
are better than GHZ states even when using time-adaptive methods.

Another resource to be taken into account is the total duration of the
experiment
$T=\sum_{i=1}^S \nu_i^{\textrm{min}}\tilde{t}_i=\sum_{i=1}^{S_1}
\nu_i^{\textrm{min}}\tilde{t}_i+\sum_{i=S_1+1}^S \tilde{t}_i$. For
measurements on an $N$-probe GHZ state, $\nu_i^\textrm{min}$ and $S_1$
are given by Eq.~\eqref{eq:mar_err} and $\tilde t_i=i/(4N)$, as
discussed above. In this case
\begin{eqnarray}
\label{eq:Tghz}
T&=&\frac{S_1}{4N}\sum_{i=1}^{S_1}\frac{i+1/2}{i}+ \frac{1}{4N}\sum_{i=S_1+1}^{S}i\nonumber\\
&\approx&\frac{1}{8N}\left[2S_1\left(S_1+\ln(S_1)-1\right)+S(S+1)\right].
\end{eqnarray}
When $S\gg 1$, using Eq.~\eqref{eq:nutot} and setting $N=1$ in
Eq.~\eqref{eq:var_method_ideal}, one can see that the bound on
$\text{Var}_{\omega}\left[ \widehat{\omega}_\textrm{MLE} \right]$
scales as $1/(N T)^{3/2}$, instead of as $1/(N T)$, which is the bound
on the error without the use of time-adaptive strategies (see
Eqs.~\eqref{eq:QCRB_totaltime_max}).

For measurements on an N-probe product state, $ \nu_i ^\textrm{min}$
is equal to the value given by Eq.~\eqref{eq:mar_err} divided by N,
$S_1\to S_1/N$ and $\tilde t_i=i/4$, leading to
\begin{eqnarray}
\label{eq:TNprobe}
T&=&\frac{S_1}{4N}\sum_{i=1}^{S_1/N}\frac{i+1/2}{i}+ \frac{1}{4}\sum_{i=S_1/N+1}^{S}i\nonumber\\
&\approx&\frac{1}{8}\left[2\frac{S_1}{N}\left(\frac{S_1}{N}+\ln(\frac{S_1}{N})-1\right)+S(S+1)\right].
\end{eqnarray}
Now, for $S\gg 1$, the use of Eq.~\eqref{eq:nutot} shows the that the
bound on
$\text{Var}_{\omega}\left[ \widehat{\omega}_\textrm{MLE} \right]$
scales as $1/(N T^{3/2})$. This scaling with $NT$ is worser than the
scaling for GHZ states. Nevertheless, it is worth to recall that,
given $T$, the number of measurements when using GHZ states is
$\sqrt N$ times the number of measurements on product states. This, in
turn, means that the total number of probes used in the estimation
with GHZ states is $\sqrt N$ times the total number of probes used in
the estimation with product states. Consequently, if $N^{3/2}$- probe
product states are used, instead of N- probe product states, the same
variance on the estimation of $\tilde \omega$ is reached, for given
$T$, as the variance obtained with N-probe GHZ states, using, in both
cases, the same total number of probes. Therefore, if the total number
of probes and the total time $T$ are counted as resources, there is no
advantage in the use of GHZ sates, when compared with the use of
product states. This will be shown numerically in section
\ref{sc:numsim}.

In order to obtain a simple analytical expression, the bound in
Eq.~\eqref{eq:var_method}, exemplified in Fig.~\ref{fig:Err_Qubit},
was derived under the assumption that the confidence levels of the
confidence intervals $CI_i$, obtained at the primary steps $i$ of the
ATFE, were equals and fixed. However, after the step $i=S_1$, the
Fisher information obtained in a single measurement is larger than the
Fisher information necessary to produce a confidence interval with
marginal error $E_i=1/(i+1)$ and confidence level $C_l$. Since the
marginal errors $E_i$ are fixed, this implies that the confidence
level $C_l$ increases after the step $i=S_1$. This, on the other side,
decreases the probability that the parameter is outside the resulting
confidence intervals, diminishing the error in the estimation of
$\omega$. This was not taken into account in deriving the bound in
Eq.~\eqref{eq:var_method}. Consequently, the actual upper bound on
$\text{Var}_{\omega}\left[ \widehat{\omega}_\textrm{MLE} \right]$ is
smaller and tighter than the bound given in that equation.
Furthermore, in deriving that bound, the confidence intervals were
updated only after $\nu_i^\textrm{min}$ measurements, inside each
primary step $i$, instead of after every measurement. During the
numerical simulations, we could see that, by updating the confidence
interval after each measurement, using the previous MLE, one obtains
smaller errors, since those estimates which were close, but outside the
confidence interval, have a chance of being inside the updated
interval. Unfortunately, we do not have an analytical expression for
the upper bound to the Holevo variance, when the two points above are
taken into account. For this reason, we shall rely on numerical
simulations for assessing the final error obtainable in the ATFE
protocol.
\subsubsection*{Extending Ramsey phase identifiability}


For product states, when individual qubits can be independently
controlled and measured, the initial phase interval where the Ramsey
phase $\phi$ is identifiable can be extended. This larger interval can
be used to extend the evolution time, $\tilde{t}$, at each adaptive
step in the ATFE protocol. One such strategy is the dual-quadrature
measurement introduced in Ref. \cite{Shaw2024}, which, by controlling
and measuring two qubits, can expand the phase interval from
$[-\pi/2, \pi/2]$ to $[-\pi, \pi]$. Furthermore, larger
extensions of the phase interval—and consequently longer evolution
times, $\tilde{t}$—can be achieved by using $N$ atoms and measuring
them over different timescales. This protocol allows the evolution
time to increase by factors of $2, \ldots, 2^{N/2-1}$m with $N$ even.

\begin{figure*}[ht]
  \centering
  \includegraphics[scale = 0.33]{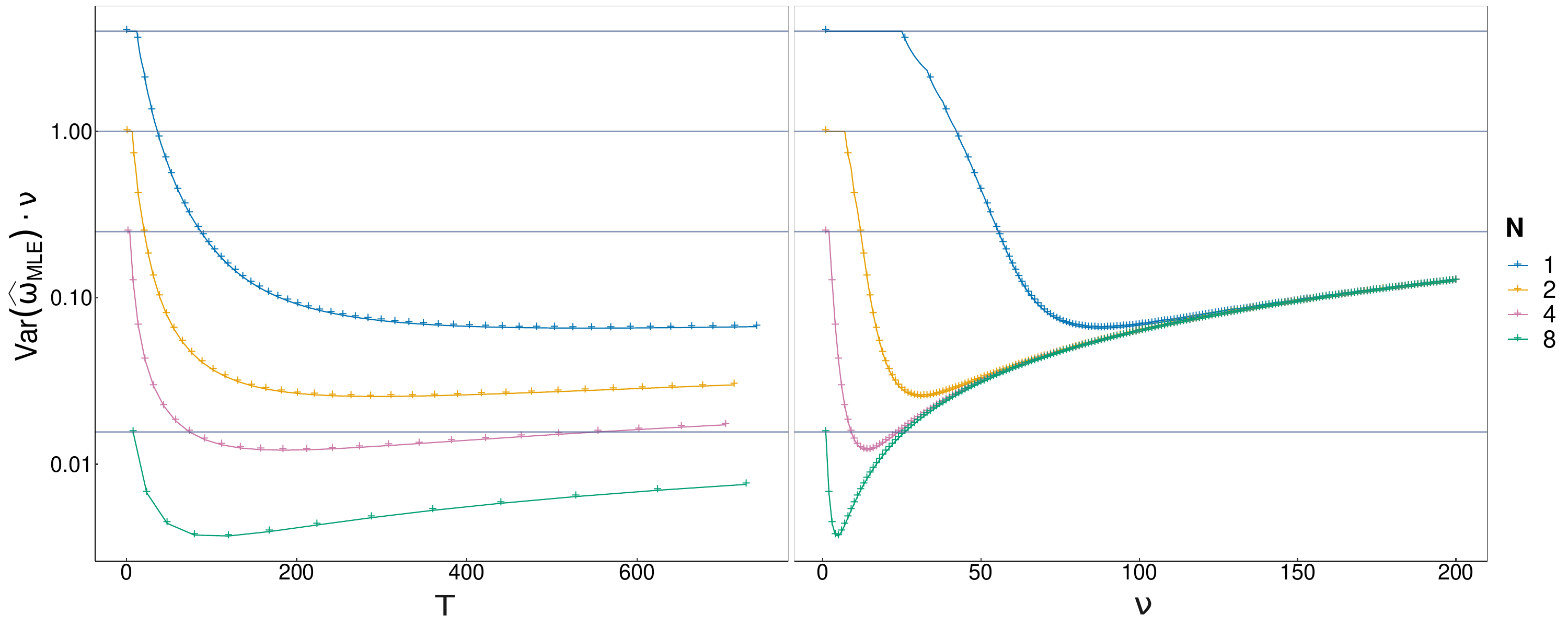}
  \caption{\label{fig:Beyond} \textbf{Variance of the MLE,
      $\text{Var}(\hat{\omega}_{\text{MLE}}) \cdot \nu$, using
      different number of atoms with the dual-quadrature measurement
      described in \cite{Shaw2024} combined with the ATFE protocol.}
    The variance is shown as a function of the total time $T$ (left)
    and the number of measurements $\nu$ (right) for different
    particle numbers, $N = {1, 2, 4, 8}$. The cases for even $N$
    correspond to the strategy of dual-quadrature measurements, while
    the curve for $N=1$ represents the ATFE with standard Ramsey
    spectroscopy, used here as a reference threshold. Both figures
    demonstrate that the ATFE protocol reduces the error compared to
    the case without adaptive measurements. As $T$ or $\nu$ increases,
    the variance decreases, reaching a minimum defined by the
    confidence interval and the initial time $\tilde{t}_1 = 1/2$ for
    $N=1$ and $\tilde{t}_1 = 2^{N/2-1}$ for even $N$. The qualitative
    behavior of the variance remains consistent for different numbers
    of qubits in both figures. The gray solid lines indicate the
    estimation error predicted by the QCRB for Ramsey spectroscopy
    when the ATFE protocol is not implemented.}
\end{figure*}

Using this technique, the maximum measurement time $t_\textrm{prod}$
in the ATFE protocol can be extended from $\pi/(2\Delta\Omega)$ to
$2^{N/2}\pi/(2\Delta\Omega)$, for $N$ even. This extension does
not alter the qualitative behavior of the error predicted by our
method; however, by allowing for larger time steps, it reduces the
error more rapidly. The trade-off is the additional complexity
required to individually control and measure each qubit. In
Figure~\ref{fig:Beyond}, we show how the ATFE protocol improves when
the dual-quadrature measurement is implemented for different numbers
of atoms.

\subsection{Numerical simulations of the protocol}\label{sc:numsim}

We summarize the ATFE strategy for GHZ states in the algorithm
\ref{algorithm_1}. For the first series of measurements, correspoding
to strategy $i=1$, each with time $\tilde{t}_1$, we choose $\nu_{1}$
in AQSE such that the asymptotic behavior of the MLE is reached (i.e.
the distribution of the estimator is approximately normal), this allow
us to use the formula Eq.~\eqref{eq:conf_intervals}.  From strategies $i>1$, we
use the minimum number of measurements, $\nu_i^\textrm{min}$, given by
Eq. (\ref{eq:mar_err}), as the distribution of the estimator is
already approximately normal since strategy $i=1$. For confidence
levels close to one (such as $0.999$), we numerically observe that the
estimator is approximately normal after $\nu_1^\textrm{min}$
measurements (see Eq.~\eqref{eq:mar_err}).
\begin{algorithm}[h!]
  \label{algorithm_1}
  \SetAlgoLined
  \KwIn{\\Confidence: $1-\alpha$, \\
    Number of particles: $N_{\textrm{GHZ}}$, \\
    Number of initial measurements: $\nu_1$,\\
    Number of total measurements: $\nu \geq \nu_1 $.}
  \KwOut{\\ Estimate $\widehat{\omega}(x)$ after $\nu$ measurements.}
  \textrm{Initialize variable: }
  $x \gets x_0 \in \left\{ 0 ,1 \right\}$\; \textrm{Initialize
    parameter space: }
  $\tilde{\Omega}_{0} = [-1, 1)$\; \textrm{First
    guess:}
  $\widehat{\omega}(x) \leftarrow
  \textrm{rand}(1,[-1,1) )$\;
  $i \gets 1$\;
  $\tilde{t}_1 \gets 1/4N_{\textrm{GHZ}}$\;
  $L(x_0;\tilde{\omega}) \gets 1$\; \For{$j = 1 \text{ to } \nu$}{

    $x_j \gets \textrm{ outcome from } P_{\textrm{L}}\left(
      \widehat{\omega}(x), \tilde{t}_{j} \right)$\;
    $L(x_j;\tilde{\omega}) \gets \textrm{Tr}\left[ P\left(x_j;
        \widehat{\omega}(x), \tilde{t}_{j} \right) \rho(\tilde{\omega}) \right]$\;
    $x \gets x || x_j$\;
    $L(x;\tilde{\omega}) \gets \prod_{x_n \in x}L(x_n;\tilde{\omega})$\;
    \eIf{ $j \geq \nu_1$ }{
      $\widehat{\omega}(x) \gets \arg\max_{\Omega_{j-1}} L(x;\tilde{\omega})$\;
    }
    {
      $\widehat{\omega}(x) \gets \arg\max_{\Omega_{0}} L(x;\tilde{\omega})$\;
    }
    $\tilde{\Omega}_{j} \gets CI(\widehat{\omega}(x); \alpha)$\;
    \eIf{ $\mathrm{E}_i(\tilde{\Omega}_j) \leq \frac{1}{i+1}$ \rm{\textbf{AND}} $j \geq \nu_{1}$ }{
      $i \gets i + 1$\;
      $\tilde{t}_{j} \gets \tilde{t}_{j-1} + 1/4N_{\textrm{GHZ}}$\;
    }
    {
      $\tilde{t}_{j+1} \gets \tilde{t}_{j}$\;
    }

    }
 \Return{$\widehat{\omega}(x)$}

  \caption{Adaptive-Time Frequency Estimation (ATFE)}
\end{algorithm}

We summarize the ATFE strategy for product states in the algorithm
\ref{algorithm_2}. In order to compare with the GHZ case, when a product state
$\rho^{\otimes N}$, consisting of $N$ qubit states is employed as a probe state,
we assume that they are measured at the same time (i.e. in parallel). Note that
in this case we have to choose the POVM that we will use in each of the $N$
qubits. For the first strategy, $i=1$, we found better results if we use a
randomly generated $\tilde{g}$ (from $-1$ to $1$) for the POVM of each qubit, as
this minimizes non-identifiability problems. Subsequently, the strategy operates
similarly to the GHZ state case, using the same POVM for each qubit.

\begin{algorithm}[h!]
  \label{algorithm_2}
  \SetAlgoLined \SetKwBlock{DoParallel}{do in parallel}{end}
  \KwIn{\\Confidence: $1-\alpha$, \\
    Size of product state: $N$, \\
    Number of initial measurements:  $\nu_1$,\\
    Number of total measurements: $\nu \geq \nu_1$.}
  \KwOut{\\ Estimate $\widehat{\omega}(x)$ after $\nu$ measurements.}
  \textrm{Initialize variable: } $x \gets x_0 \in \left\{ 0 ,1 \right\}$\;
  \textrm{Initialize parameter space: } $\tilde{\Omega}_{0} = [-1, 1)$\; \textrm{First
    parameters for the POVMs:} $\tilde{g} \in \mathbb{R}^{N} \leftarrow
  \textrm{rand}(N, [-1,1 ) )$\;
  $i \gets 1$\; $\tilde{t}_1 \gets
  1/4$\;
  $L(x_0;\tilde{\omega}) \gets 1$\;
  \For{$j = 1 \text{ to } \nu$}{
    \DoParallel{ \For{$ 1 \leq k \leq N$}{ $x_k \gets \textrm{ outcome from }
        P_{\textrm{L}}\left(\tilde{g}[k], \tilde{t}_j \right)$\;
        $L(x_k;\tilde{\omega}) \gets \textrm{Tr}\left[ P\left(x_k; \tilde{g}[k], \tilde{t}_j \right)
          \rho(\tilde{\omega}) \right]$\; } } $x \gets (x_1,x_2,\ldots,x_N)
    \mathbin\Vert x$\;
    $L(x;\tilde{\omega}) \gets \prod_{x_n \in x} L(x_n; \tilde{\omega})$\;
    \eIf{ $j \geq \nu_1$ }{ $\widehat{\omega}(x) \gets
      \arg\max_{\tilde{\Omega}_{j-1}} L(x;\tilde{\omega})$\; } { $\widehat{\omega}(x)
      \gets \arg\max_{\tilde{\Omega}_{0}} L(x;\tilde{\omega})$\; } $\tilde{g} \leftarrow
    \textrm{rep}\left(\widehat{\omega}(x), N\right)$\;
    $\Omega_{j} \gets
    CI(\widehat{\omega}(x); \alpha)$\;
    \eIf{ $\mathrm{E}_i(\tilde{\Omega}_j) \leq
      \frac{1}{i+1}$ \rm{\textbf{AND}} $j \geq \nu_1$ }{ $i \gets i + 1$\;
      $\tilde{t}_{j+1} \gets \tilde{t}_{j} + 1/4$\; } { $\tilde{t}_{j+1} \gets
      \tilde{t}_{j}$\;
  }
  }
  \Return{$\widehat{\omega}(x)$}
 \caption{Adaptive-Time Frequency Estimation (ATFE) in Parallel}
\end{algorithm}

We now present the results obtained from the simulation of the algorithms for
ATFE. First we focus on the case of $N=1$ qubit, (in this case
algorithm~\ref{algorithm_1} or~\ref{algorithm_2} are the same). The error as a
function of the number of measurements is shown in
Fig.~\ref{fig:Numerical_Qubit}. The solid gray line represents the estimation
error predicted by the QCRB for $\tilde{t} = 1/4$, which corresponds to the case
without adaptive measurements; an estimation error below that line shows the
advantage of the adaptive method. The dark blue rhombus represent the lower
bound for the estimation error of the ATFE strategy without update of the
confidence interval center (see Eq.~\eqref{eq:var_method}). The green points
corresponds to the numerical simulation of that case, which shows that the QCRB
for the non-time-adaptive strategy ($\tilde{t}=1/4$) can be improved. As
predicted by Eq.~\eqref{eq:var_method}, for a large number of measurements
($\nu$), the contribution to the error due to the estimations outside the
confidence intervals will eventually dominate. The light blue rhombus in Fig.
\ref{fig:Numerical_Qubit} represent the smallest possible error, which is the
QCRB ($C_l=1$). Our goal is to achieve an estimation error that is as close as
possible to this bound. The results obtained when the confidence interval is
updated at each step of the simulation are shown by the red points in Fig.
\ref{fig:Numerical_Qubit}. The error is found to be close to the smallest
possible error for $\nu$ up to $60$, indicating the large advantage of the
proposed measurement strategy. For the case of product states, increasing the
number of qubits diminishes the error by a factor of $N$, as predicted by the
QCRB.

\begin{figure}[h!]
   \centering \includegraphics[scale = 0.33]{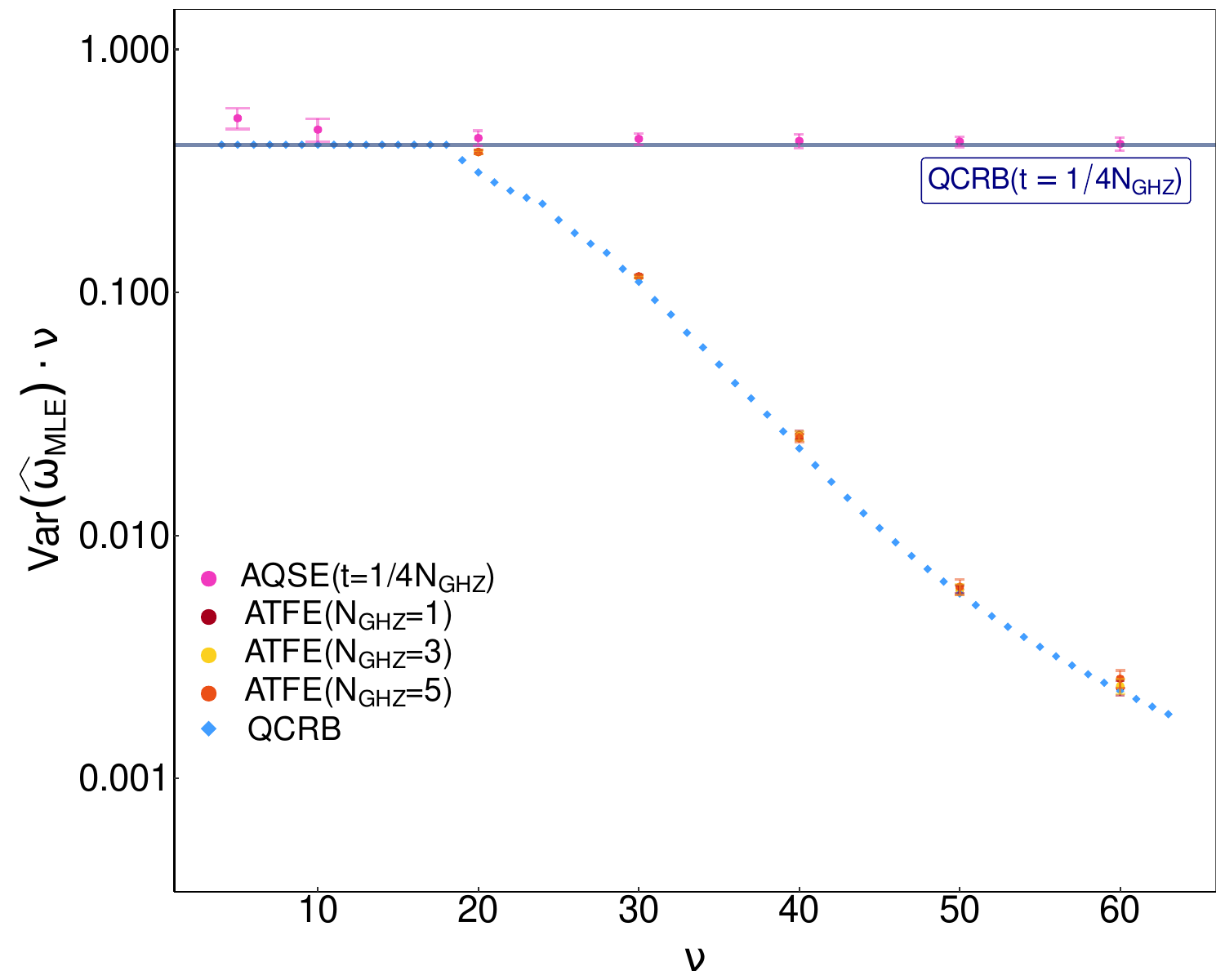}
   \caption{\textbf{Holevo variance vs the number of measurements for different
       $N_{\rm{GHZ}}$}. The error of ATFE, with update of the confidence
     interval at each step, for different number of particles $N_{\rm{GHZ}} =
     1,5,10$ and initial time $\tilde{t}_1 = 1/(4N_{\rm{GHZ}})$. Increasing the
     number of particles in the GHZ state does not improve the error as a
     function of the number of performed experiments $\nu$. The error of the
     AQSE method for $\tilde{t}=1/(4N_{\rm{GHZ}})$ (gray line) and for the ideal
     case of $C_l=1$ (dark blue dots) are shown as reference. The expected
     Holevo variance at each point for each ATFE strategies was obtained from
     the average of 5 Monte Carlo simulations of the experiment, each with
     $10^3$ samples, }
  \label{fig:Numerical_Qubit_2}
\end{figure}

We now study what happens when  GHZ states are used as probes. Fig.
\ref{fig:Numerical_Qubit_2} shows the error in frequency estimation
using the algorithm \ref{algorithm_1} for ATFE with
$N_{\rm{GHZ}} = 1, 5 ,10$ and $C_l = 0.999$. Again, the solid green
line  represents the estimation error
predicted by the QCRB for $\tilde{t} = 1/4$, and the dark blue rhombus
 represent the smallest possible
error, which is the QCRB ($C_l=1$). The pink points show the error of
AQSE for $\tilde{t}=1/4$, the red, light green, and orange points the
estimation error for ATFE with $N_{\rm{GHZ}} = 1, 3,5$ respectively.
As we anticipated from Eq.~\eqref{eq:noaqsescale} and from
Eq.~\eqref{eq:var_method_ideal} with $N=1$, the estimation error does
not improve by increasing the number of qubits in a GHZ state.

\subsection{Product states vs GHZ states}

By using the maximum time that ensures an identifiable estimator at any
measurement, one obtains larger errors when GHZ states are the probe states
instead of product states. This fact can be seen by comparing equation
Eq.~\eqref{eq:QCRB_frequency_max_ramsey} with equation
Eq.~\eqref{eq:noaqsescale}, where it is clear that, at each measurement, one
learns more about the parameter using product states than with GHZ states. This
advantage remains even when time-adaptive strategies are employed. The bound
given in Eq.~\eqref{eq:var_method_ideal}, obtained without updating the
$\widehat{CI}$ at each measurement step, confirms this prediction, since the use
of initial GHZ states with arbitrary number of probes corresponds to putting
$N=1$ in that bound. Numerical simulations of the frequency estimation, where
the confidence interval is updated at subsequent adaptive steps (with the
updating process taking place after the $\nu_1$ initial steps), further validate
that GHZ states do not outperform product states when considering the total
number of qubits used in the experiment (see Fig.~\ref{fig:Numerical_Qubit_2}).

\begin{figure}[h!]
  \centering \includegraphics[scale = 0.3]{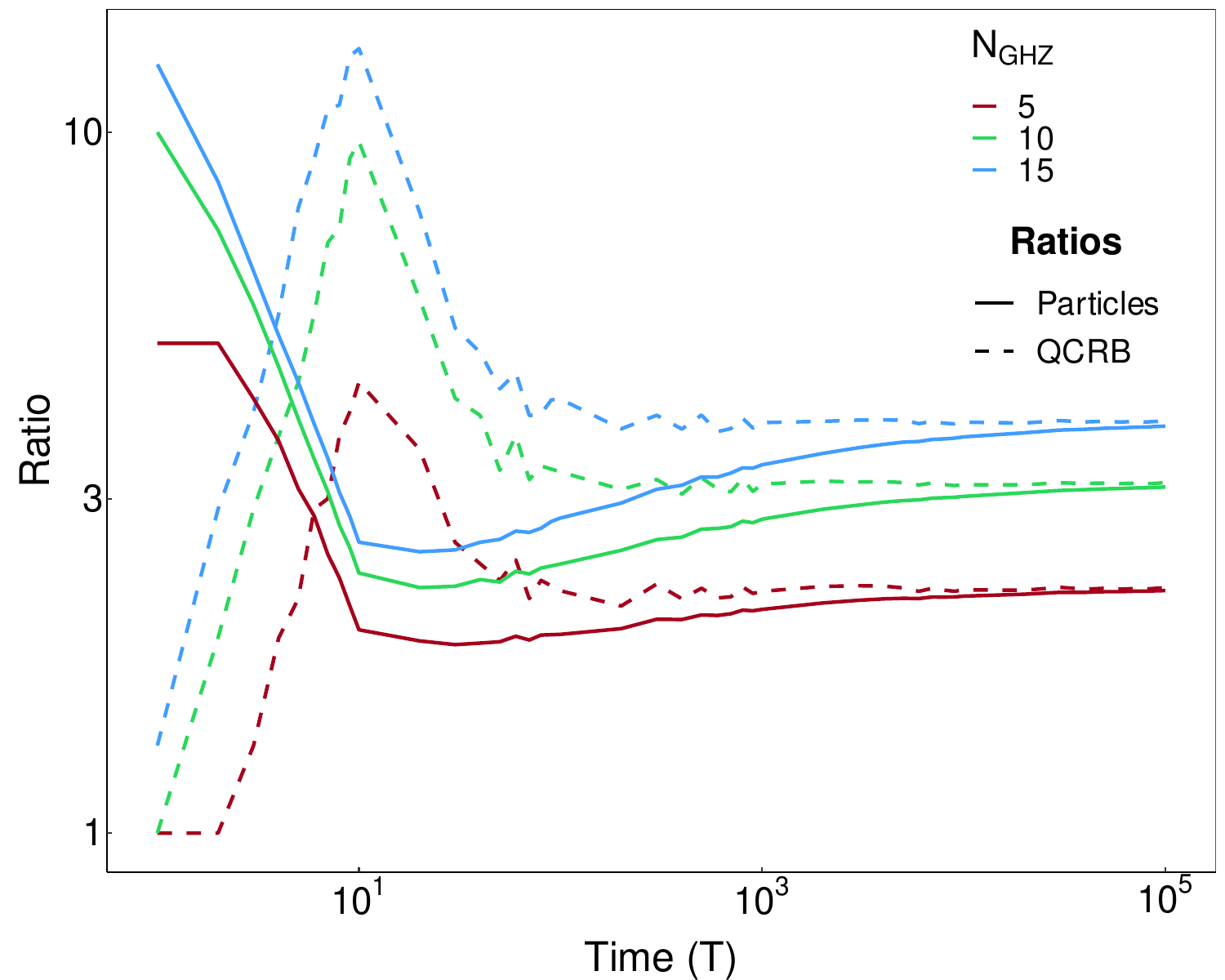}
  \caption{ \textbf{Comparison of total number of qubits and variance for GHZ
      and product states in parallel for ATFE at $0.999\%$ of confidence.} The
    ratios between the total number of particles and the corresponding QCRBs are
    examined as a function of time. The comparison encompasses the utilization
    of GHZ states and product states in parallel for ATFE. Each color
    corresponds to a distinct dimension of the Hilbert space, with red
    representing 5 dimensions, green representing 10 dimensions, and blue
    representing 15 dimensions. The solid line illustrates the ratio of the
    number of particles, while the dashed line depicts the ratio of variances. }
  \label{fig:Numerical_GHZvsProd}
\end{figure}

On the other hand, when considering the minimization of the estimation error
with respect to the total experiment time $T$, the ATFE protocol using GHZ
states exhibits smaller estimator variance compared to product states. This
behavior is in accordance with the predictions given by
Eqs.~\eqref{eq:var_method_ideal}, \eqref{eq:Tghz} and \eqref{eq:TNprobe}, in the
case where the confidence interval is not updated at each increment of time.
When the confidence interval is updated at each adaptive step, the GHZ states
give smaller errors than product states, confirming those predictions. This is
shown in Fig.~\ref{fig:Numerical_GHZvsProd}, where the ratio between the
estimator variance when the probe is a $N$-product state and the estimator
variance for a GHZ state with $N$ qubits is plotted. This ratio scales as
$\sqrt{N}$ when the total time $T$ is large enough, as predict by
Eqs.~\eqref{eq:Tghz} and \eqref{eq:TNprobe}. However, as already discussed, for
a fixed time $T$, the total number of qubits employed using GHZ states is larger
by $\sqrt{N}$ than the total number of qubits employed with product states, as
more measurements are needed with GHZ states. In
Fig.~\ref{fig:Numerical_GHZvsProd} we also plot the ratio between the total
number of probes used for GHZ states and for product states. Specifically, we
explore scenarios involving $N_{\rm{GHZ}} = 5, 10, 15$ for ATFE with GHZ states,
as well as a set of $N = 5, 10, 15$ product states $\rho(\omega)^{\otimes N}$,
which are measured in parallel at each adaptive step. In every case, we maintain
a confidence level of $C_l = 0.999$ and set $\nu_1 = 20$ to ensure asymptotic
normality in the distribution of the MLE after the first strategy ($i=1$). It
can be seen that the ratio between the variances tends to the same value as the
ratio between the number of particles. This means that $\sqrt{N}$ more qubits
were used with GHZ states than with product states. On the other side, since
increasing the total number of probes by $\sqrt{N}$, for product states,
decreases the estimation variance by the same factor, this implies that, when
the total number $N_T$ of probes and the total time $T$ of the estimation are
fixed, product states and GHZ have the same performance for the time-adaptive
strategy presented here, in contrast to the non-adaptive strategy, where the use
of product states is more advantageous than the use of GHZ states.

\section{Discussion \& Conclusions}
In this article, we investigated the real advantages of the use of entangled
states for improving frequency estimation, with special focus on the
spectroscopy of two-level systems. The basic metrological scheme to estimating a
transition frequency $\omega$, using $N$ two-level probes, consists in first
preparing the $N$ probes in an adequate quantum state, letting them evolve for a
time $t$ and then measuring some observable on the final state. The measurement
results are fed into an estimator function, which produces an estimate of
$\omega$. The QCRB provides the ultimate lower bound on the uncertainty of the
estimate of the value of $\omega$, and in principle, this bound can be reached
using the maximum likelihood estimator. That bound shows that, when the
estimation procedure consists of a large number of preparation-sensing cycles,
with a fixed sensing time and a fixed number of probes in each run of the
experiment, the use of probes in a maximally entangled state leads to an
enhanced precision in the frequency estimation when compared with the use of the
same number of probes in a product state. However, in order to that bound be
reachable it is necessary that the statistical model be identifiable. This, in
turn, puts a limit on the maximal sensing time allowed in each run of the
estimation procedure. We have discussed in detail the effects of this
restriction on the maximal sensing time on the de facto reachable bound on the
precision of the frequency estimation and, in particular, how the requirement on
the identifiability produces a maximum likelihood estimator that annuls any
possible advantage of the use of maximally entangled probes when compared with
use of independent probes. In fact, when the total number of probes and the
total sensing time are counted as resources, it is more advantageous to use
independent probes than maximally entangled ones.

As a means to counteract the limitations on the maximal sensing time in
frequency estimation, imposed by the requirement of statistical model
identifiability, we presented a time-adaptive estimation strategy. In this
strategy, one assumes that some prior knowledge about the value of the frequency
$\omega$ already exists and that this value lies inside some frequency interval
of a given length. Using the maximal sensing time allowed by that interval
length, one starts with a small number of identical
preparation-sensing-measurement cycles to produce a first estimation of
$\omega$. This first estimation allows one to shorten the interval length and,
consequently, to increase the sensing time for the next estimation cycle. This
process is repeated until the desired estimation uncertainty is reached. We have
determined a reachable error bound for the presented strategy and discussed how
to optimally choose its parameters in order to minimize that bound. The bound
shows that the time-adaptive strategy leads to much better scaling of the
estimation uncertainty with the total number of probes and the total sensing
time than the traditional fixed-sensing-time strategy. The bound also shows
that, when the total number of probes and the total sensing time are counted as
resources, independent probes and maximally entangled ones have now the same
performance, in contrast to the non-adaptive strategy, where the use of product
states is more advantageous than the use of maximally entangled states. Finally,
we presented numerical simulations of the proposed time-adaptive estimation
strategy. Those simulations confirmed all the analytical predictions presented
in this article.

\section*{Acknowledgments}
We thank Laboratorio Universitario de Cómputo de Alto Rendimiento (LUCAR) of
IIMAS-UNAM for their service on information processing. M.R.-G. acknowledges
funding from the National Science Foundation (NSF) Grants FRHTP No. PHY-2116246
and No. QLCI-2016244. P.B.-B. gratefully acknowledges support from the
PAPIIT-DGAPA Grant No. IG101324. R.L.M.F. acknowledges the support of the John
Templeton Foundation (Grant No. 62424).



\begin{thebibliography}{40}%
\makeatletter
\providecommand \@ifxundefined [1]{%
 \@ifx{#1\undefined}
}%
\providecommand \@ifnum [1]{%
 \ifnum #1\expandafter \@firstoftwo
 \else \expandafter \@secondoftwo
 \fi
}%
\providecommand \@ifx [1]{%
 \ifx #1\expandafter \@firstoftwo
 \else \expandafter \@secondoftwo
 \fi
}%
\providecommand \natexlab [1]{#1}%
\providecommand \enquote  [1]{``#1''}%
\providecommand \bibnamefont  [1]{#1}%
\providecommand \bibfnamefont [1]{#1}%
\providecommand \citenamefont [1]{#1}%
\providecommand \href@noop [0]{\@secondoftwo}%
\providecommand \href [0]{\begingroup \@sanitize@url \@href}%
\providecommand \@href[1]{\@@startlink{#1}\@@href}%
\providecommand \@@href[1]{\endgroup#1\@@endlink}%
\providecommand \@sanitize@url [0]{\catcode `\\12\catcode `\$12\catcode
  `\&12\catcode `\#12\catcode `\^12\catcode `\_12\catcode `\%12\relax}%
\providecommand \@@startlink[1]{}%
\providecommand \@@endlink[0]{}%
\providecommand \url  [0]{\begingroup\@sanitize@url \@url }%
\providecommand \@url [1]{\endgroup\@href {#1}{\urlprefix }}%
\providecommand \urlprefix  [0]{URL }%
\providecommand \Eprint [0]{\href }%
\providecommand \doibase [0]{http://dx.doi.org/}%
\providecommand \selectlanguage [0]{\@gobble}%
\providecommand \bibinfo  [0]{\@secondoftwo}%
\providecommand \bibfield  [0]{\@secondoftwo}%
\providecommand \translation [1]{[#1]}%
\providecommand \BibitemOpen [0]{}%
\providecommand \bibitemStop [0]{}%
\providecommand \bibitemNoStop [0]{.\EOS\space}%
\providecommand \EOS [0]{\spacefactor3000\relax}%
\providecommand \BibitemShut  [1]{\csname bibitem#1\endcsname}%
\let\auto@bib@innerbib\@empty
\bibitem [{\citenamefont {Godun}\ \emph {et~al.}(2014)\citenamefont {Godun},
  \citenamefont {Nisbet-Jones}, \citenamefont {Jones}, \citenamefont {King},
  \citenamefont {Johnson}, \citenamefont {Margolis}, \citenamefont {Szymaniec},
  \citenamefont {Lea}, \citenamefont {Bongs},\ and\ \citenamefont
  {Gill}}]{PhysRevLett.113.210801}%
  \BibitemOpen
  \bibfield  {author} {\bibinfo {author} {\bibfnamefont {R.~M.}\ \bibnamefont
  {Godun}}, \bibinfo {author} {\bibfnamefont {P.~B.~R.}\ \bibnamefont
  {Nisbet-Jones}}, \bibinfo {author} {\bibfnamefont {J.~M.}\ \bibnamefont
  {Jones}}, \bibinfo {author} {\bibfnamefont {S.~A.}\ \bibnamefont {King}},
  \bibinfo {author} {\bibfnamefont {L.~A.~M.}\ \bibnamefont {Johnson}},
  \bibinfo {author} {\bibfnamefont {H.~S.}\ \bibnamefont {Margolis}}, \bibinfo
  {author} {\bibfnamefont {K.}~\bibnamefont {Szymaniec}}, \bibinfo {author}
  {\bibfnamefont {S.~N.}\ \bibnamefont {Lea}}, \bibinfo {author} {\bibfnamefont
  {K.}~\bibnamefont {Bongs}}, \ and\ \bibinfo {author} {\bibfnamefont
  {P.}~\bibnamefont {Gill}},\ }\bibfield  {title} {\emph {\enquote {\bibinfo
  {title} {Frequency ratio of two optical clock transitions in
  $^{171}{\mathrm{yb}}^{+}$ and constraints on the time variation of
  fundamental constants},}\ }}\href {\doibase 10.1103/PhysRevLett.113.210801}
  {\bibfield  {journal} {\bibinfo  {journal} {Phys. Rev. Lett.}\ }\textbf
  {\bibinfo {volume} {113}},\ \bibinfo {pages} {210801} (\bibinfo {year}
  {2014})}\BibitemShut {NoStop}%
\bibitem [{\citenamefont {Arvanitaki}\ \emph {et~al.}(2015)\citenamefont
    {Arvanitaki}, \citenamefont {Huang},\ and\ \citenamefont
    {Van~Tilburg}}]{PhysRevD.91.015015}%
  \BibitemOpen \bibfield {author} {\bibinfo {author} {\bibfnamefont
      {A.}~\bibnamefont {Arvanitaki}}, \bibinfo {author} {\bibfnamefont
      {J.}~\bibnamefont {Huang}}, \ and\ \bibinfo {author} {\bibfnamefont
      {K.}~\bibnamefont {Van~Tilburg}},\ }\bibfield {title} {\emph {\enquote
      {\bibinfo {title} {Searching for dilaton dark matter with atomic
          clocks},}\ }}\href {\doibase 10.1103/PhysRevD.91.015015}
  {\bibfield {journal} {\bibinfo {journal} {Phys. Rev. D}\ }\textbf {\bibinfo
      {volume} {91}},\ \bibinfo {pages} {015015} (\bibinfo {year}
    {2015})}\BibitemShut {NoStop}%
\bibitem [{\citenamefont {Barontini}\ \emph {et~al.}(2022)\citenamefont
    {Barontini}, \citenamefont {Blackburn}, \citenamefont {Boyer}, \citenamefont
    {Butuc-Mayer}, \citenamefont {Calmet}, \citenamefont {Crespo López-Urrutia},
    \citenamefont {Curtis}, \citenamefont {Darquié}, \citenamefont {Dunningham},
    \citenamefont {Fitch}, \citenamefont {Forgan}, \citenamefont {Georgiou},
    \citenamefont {Gill}, \citenamefont {Godun}, \citenamefont {Goldwin},
    \citenamefont {Guarrera}, \citenamefont {Harwood}, \citenamefont {Hill},
    \citenamefont {Hendricks}, \citenamefont {Jeong}, \citenamefont {Johnson},
    \citenamefont {Keller}, \citenamefont {Kozhiparambil~Sajith}, \citenamefont
    {Kuipers}, \citenamefont {Margolis}, \citenamefont {Mayo}, \citenamefont
    {Newman}, \citenamefont {Parsons}, \citenamefont {Prokhorov}, \citenamefont
    {Robertson}, \citenamefont {Rodewald}, \citenamefont {Safronova},
    \citenamefont {Sauer}, \citenamefont {Schioppo}, \citenamefont {Sherrill},
    \citenamefont {Stadnik}, \citenamefont {Szymaniec}, \citenamefont {Tarbutt},
    \citenamefont {Thompson}, \citenamefont {Tofful}, \citenamefont {Tunesi},
    \citenamefont {Vecchio}, \citenamefont {Wang},\ and\ \citenamefont
    {Worm}}]{Barontini2022}%
  \BibitemOpen \bibfield {author} {\bibinfo {author} {\bibfnamefont
      {G.}~\bibnamefont {Barontini}}, \bibinfo {author} {\bibfnamefont
      {L.}~\bibnamefont {Blackburn}}, \bibinfo {author} {\bibfnamefont
      {V.}~\bibnamefont {Boyer}}, \bibinfo {author} {\bibfnamefont
      {F.}~\bibnamefont {Butuc-Mayer}}, \bibinfo {author} {\bibfnamefont
      {X.}~\bibnamefont {Calmet}}, \bibinfo {author} {\bibfnamefont {J.~R.}\
      \bibnamefont {Crespo López-Urrutia}}, \bibinfo {author} {\bibfnamefont
      {E.~A.}\ \bibnamefont {Curtis}}, \bibinfo {author} {\bibfnamefont
      {B.}~\bibnamefont {Darquié}}, \bibinfo {author} {\bibfnamefont
      {J.}~\bibnamefont {Dunningham}}, \emph {et~al.},\ }\bibfield {title}
  {\emph {\enquote {\bibinfo {title} {Measuring the stability of fundamental
          constants with a network of clocks},}\ }}\href
  {\doibase 10.1140/epjqt/s40507-022-00130-5} {\bibfield {journal} {\bibinfo
      {journal} {EPJ Quantum Technology}\ }\textbf {\bibinfo {volume} {9}}
    (\bibinfo {year} {2022}),\ 10.1140/epjqt/s40507-022-00130-5}\BibitemShut
  {NoStop}%
\bibitem [{\citenamefont {Danilin}\ \emph {et~al.}(2018)\citenamefont
  {Danilin}, \citenamefont {Lebedev}, \citenamefont {Veps\"{a}l\"{a}inen},
  \citenamefont {Lesovik}, \citenamefont {Blatter},\ and\ \citenamefont
  {Paraoanu}}]{danilin2018quantum}%
  \BibitemOpen
  \bibfield  {author} {\bibinfo {author} {\bibfnamefont {S.}~\bibnamefont
  {Danilin}}, \bibinfo {author} {\bibfnamefont {A.~V.}\ \bibnamefont
  {Lebedev}}, \bibinfo {author} {\bibfnamefont {A.}~\bibnamefont
  {Veps\"{a}l\"{a}inen}}, \bibinfo {author} {\bibfnamefont {G.~B.}\
  \bibnamefont {Lesovik}}, \bibinfo {author} {\bibfnamefont {G.}~\bibnamefont
  {Blatter}}, \ and\ \bibinfo {author} {\bibfnamefont {G.~S.}\ \bibnamefont
  {Paraoanu}},\ }\bibfield  {title} {\emph {\enquote {\bibinfo {title}
  {Quantum-enhanced magnetometry by phase estimation algorithms with a single
  artificial atom},}\ }}\href {\doibase 10.1038/s41534-018-0078-y} {\bibfield
  {journal} {\bibinfo  {journal} {npj Quantum Information}\ }\textbf {\bibinfo
  {volume} {4}} (\bibinfo {year} {2018}),\
  10.1038/s41534-018-0078-y}\BibitemShut {NoStop}%
\bibitem [{\citenamefont {Dong}\ \emph {et~al.}(2021)\citenamefont {Dong},
    \citenamefont {Xue}, \citenamefont {Luo}, \citenamefont {Ge}, \citenamefont
    {Liu}, \citenamefont {Yuan}, \citenamefont {Zhang},\ and\ \citenamefont
    {Zhu}}]{dong2021high}%
  \BibitemOpen \bibfield {author} {\bibinfo {author} {\bibfnamefont
      {H.}~\bibnamefont {Dong}}, \bibinfo {author} {\bibfnamefont
      {L.}~\bibnamefont {Xue}}, \bibinfo {author} {\bibfnamefont
      {W.}~\bibnamefont {Luo}}, \bibinfo {author} {\bibfnamefont
      {J.}~\bibnamefont {Ge}}, \bibinfo {author} {\bibfnamefont
      {H.}~\bibnamefont {Liu}}, \bibinfo {author} {\bibfnamefont
      {Z.}~\bibnamefont {Yuan}}, \bibinfo {author} {\bibfnamefont
      {H.}~\bibnamefont {Zhang}}, \ and\ \bibinfo {author} {\bibfnamefont
      {J.}~\bibnamefont {Zhu}},\ }\bibfield {title} {\emph {\enquote {\bibinfo
        {title} {A high-accuracy and non-intermittent frequency measurement
          method for larmor signal of optically pumped cesium magnetometer},}\
    }}\href {\doibase 10.1088/1748-0221/16/06/P06001}
  {\bibfield {journal} {\bibinfo {journal} {Journal of Instrumentation}\
    }\textbf {\bibinfo {volume} {16}},\ \bibinfo {pages} {P06001} (\bibinfo
    {year} {2021})}\BibitemShut {NoStop}%
\bibitem [{\citenamefont {Wu}\ \emph {et~al.}(2019)\citenamefont {Wu},
  \citenamefont {Pagel}, \citenamefont {Malek}, \citenamefont {Nguyen},
  \citenamefont {Zi}, \citenamefont {Scheirer},\ and\ \citenamefont
  {M\"{u}ller}}]{wu2019gravity}%
  \BibitemOpen
  \bibfield  {author} {\bibinfo {author} {\bibfnamefont {X.}~\bibnamefont
  {Wu}}, \bibinfo {author} {\bibfnamefont {Z.}~\bibnamefont {Pagel}}, \bibinfo
  {author} {\bibfnamefont {B.~S.}\ \bibnamefont {Malek}}, \bibinfo {author}
  {\bibfnamefont {T.~H.}\ \bibnamefont {Nguyen}}, \bibinfo {author}
  {\bibfnamefont {F.}~\bibnamefont {Zi}}, \bibinfo {author} {\bibfnamefont
  {D.~S.}\ \bibnamefont {Scheirer}}, \ and\ \bibinfo {author} {\bibfnamefont
  {H.}~\bibnamefont {M\"{u}ller}},\ }\bibfield  {title} {\emph {\enquote
  {\bibinfo {title} {Gravity surveys using a mobile atom interferometer},}\
  }}\href {\doibase 10.1126/sciadv.aax0800} {\bibfield  {journal} {\bibinfo
  {journal} {Science Advances}\ }\textbf {\bibinfo {volume} {5}} (\bibinfo
  {year} {2019}),\ 10.1126/sciadv.aax0800}\BibitemShut {NoStop}%
\bibitem [{\citenamefont {Stray}\ \emph {et~al.}(2022)\citenamefont {Stray},
  \citenamefont {Lamb}, \citenamefont {Kaushik}, \citenamefont {Vovrosh},
  \citenamefont {Rodgers}, \citenamefont {Winch}, \citenamefont {Hayati},
  \citenamefont {Boddice}, \citenamefont {Stabrawa}, \citenamefont {Niggebaum}
  \emph {et~al.}}]{stray2022quantum}%
  \BibitemOpen
  \bibfield  {author} {\bibinfo {author} {\bibfnamefont {B.}~\bibnamefont
  {Stray}}, \bibinfo {author} {\bibfnamefont {A.}~\bibnamefont {Lamb}},
  \bibinfo {author} {\bibfnamefont {A.}~\bibnamefont {Kaushik}}, \bibinfo
  {author} {\bibfnamefont {J.}~\bibnamefont {Vovrosh}}, \bibinfo {author}
  {\bibfnamefont {A.}~\bibnamefont {Rodgers}}, \bibinfo {author} {\bibfnamefont
  {J.}~\bibnamefont {Winch}}, \bibinfo {author} {\bibfnamefont
  {F.}~\bibnamefont {Hayati}}, \bibinfo {author} {\bibfnamefont
  {D.}~\bibnamefont {Boddice}}, \bibinfo {author} {\bibfnamefont
  {A.}~\bibnamefont {Stabrawa}},  \emph {et~al.},\ }\bibfield  {title} {\emph
  {\enquote {\bibinfo {title} {Quantum sensing for gravity cartography},}\
  }}\href {\doibase 10.1038/s41586-021-04315-3} {\bibfield  {journal} {\bibinfo
   {journal} {Nature}\ }\textbf {\bibinfo {volume} {602}},\ \bibinfo {pages}
  {590} (\bibinfo {year} {2022})}\BibitemShut {NoStop}%
\bibitem [{\citenamefont {Macieszczak}\ \emph {et~al.}(2014)\citenamefont
  {Macieszczak}, \citenamefont {Fraas},\ and\ \citenamefont
  {Demkowicz-Dobrza{\'n}ski}}]{macieszczak2014bayesian}%
  \BibitemOpen
  \bibfield  {author} {\bibinfo {author} {\bibfnamefont {K.}~\bibnamefont
  {Macieszczak}}, \bibinfo {author} {\bibfnamefont {M.}~\bibnamefont {Fraas}},
  \ and\ \bibinfo {author} {\bibfnamefont {R.}~\bibnamefont
  {Demkowicz-Dobrza{\'n}ski}},\ }\bibfield  {title} {\emph {\enquote {\bibinfo
  {title} {Bayesian quantum frequency estimation in presence of collective
  dephasing},}\ }}\href {\doibase 10.1088/1367-2630/16/11/113002} {\bibfield
  {journal} {\bibinfo  {journal} {New Journal of Physics}\ }\textbf {\bibinfo
  {volume} {16}},\ \bibinfo {pages} {113002} (\bibinfo {year}
  {2014})}\BibitemShut {NoStop}%
\bibitem [{\citenamefont {Sanner}\ \emph {et~al.}(2019)\citenamefont {Sanner},
  \citenamefont {Huntemann}, \citenamefont {Lange}, \citenamefont {Tamm},
  \citenamefont {Peik}, \citenamefont {Safronova},\ and\ \citenamefont
  {Porsev}}]{sanner2019optical}%
  \BibitemOpen
  \bibfield  {author} {\bibinfo {author} {\bibfnamefont {C.}~\bibnamefont
  {Sanner}}, \bibinfo {author} {\bibfnamefont {N.}~\bibnamefont {Huntemann}},
  \bibinfo {author} {\bibfnamefont {R.}~\bibnamefont {Lange}}, \bibinfo
  {author} {\bibfnamefont {C.}~\bibnamefont {Tamm}}, \bibinfo {author}
  {\bibfnamefont {E.}~\bibnamefont {Peik}}, \bibinfo {author} {\bibfnamefont
  {M.~S.}\ \bibnamefont {Safronova}}, \ and\ \bibinfo {author} {\bibfnamefont
  {S.~G.}\ \bibnamefont {Porsev}},\ }\bibfield  {title} {\emph {\enquote
  {\bibinfo {title} {Optical clock comparison for lorentz symmetry testing},}\
  }}\href {\doibase 10.1038/s41586-019-0972-2} {\bibfield  {journal} {\bibinfo
  {journal} {Nature}\ }\textbf {\bibinfo {volume} {567}},\ \bibinfo {pages}
  {204} (\bibinfo {year} {2019})}\BibitemShut {NoStop}%
\bibitem [{\citenamefont {Madjarov}\ \emph {et~al.}(2019)\citenamefont
  {Madjarov}, \citenamefont {Cooper}, \citenamefont {Shaw}, \citenamefont
  {Covey}, \citenamefont {Schkolnik}, \citenamefont {Yoon}, \citenamefont
  {Williams},\ and\ \citenamefont {Endres}}]{PhysRevX.9.041052}%
  \BibitemOpen
  \bibfield  {author} {\bibinfo {author} {\bibfnamefont {I.~S.}\ \bibnamefont
  {Madjarov}}, \bibinfo {author} {\bibfnamefont {A.}~\bibnamefont {Cooper}},
  \bibinfo {author} {\bibfnamefont {A.~L.}\ \bibnamefont {Shaw}}, \bibinfo
  {author} {\bibfnamefont {J.~P.}\ \bibnamefont {Covey}}, \bibinfo {author}
  {\bibfnamefont {V.}~\bibnamefont {Schkolnik}}, \bibinfo {author}
  {\bibfnamefont {T.~H.}\ \bibnamefont {Yoon}}, \bibinfo {author}
  {\bibfnamefont {J.~R.}\ \bibnamefont {Williams}}, \ and\ \bibinfo {author}
  {\bibfnamefont {M.}~\bibnamefont {Endres}},\ }\bibfield  {title} {\emph
  {\enquote {\bibinfo {title} {An atomic-array optical clock with single-atom
  readout},}\ }}\href {\doibase 10.1103/PhysRevX.9.041052} {\bibfield
  {journal} {\bibinfo  {journal} {Phys. Rev. X}\ }\textbf {\bibinfo {volume}
  {9}},\ \bibinfo {pages} {041052} (\bibinfo {year} {2019})}\BibitemShut
  {NoStop}%
\bibitem [{\citenamefont {Braunstein}\ and\ \citenamefont
  {Caves}(1994)}]{Braunstein1994}%
  \BibitemOpen
  \bibfield  {author} {\bibinfo {author} {\bibfnamefont {S.~L.}\ \bibnamefont
  {Braunstein}}\ and\ \bibinfo {author} {\bibfnamefont {C.~M.}\ \bibnamefont
  {Caves}},\ }\bibfield  {title} {\emph {\enquote {\bibinfo {title}
  {Statistical distance and the geometry of quantum states},}\ }}\href
  {\doibase 10.1103/PhysRevLett.72.3439} {\bibfield  {journal} {\bibinfo
  {journal} {Phys. Rev. Lett.}\ }\textbf {\bibinfo {volume} {72}},\ \bibinfo
  {pages} {3439} (\bibinfo {year} {1994})}\BibitemShut {NoStop}%
\bibitem [{\citenamefont {Bollinger}\ \emph {et~al.}(1996)\citenamefont
    {Bollinger}, \citenamefont {Itano}, \citenamefont {Wineland},\ and\
    \citenamefont {Heinzen}}]{Bollinger}%
  \BibitemOpen \bibfield {author} {\bibinfo {author} {\bibfnamefont {J.~J.~.}\
      \bibnamefont {Bollinger}}, \bibinfo {author} {\bibfnamefont {W.~M.}\
      \bibnamefont {Itano}}, \bibinfo {author} {\bibfnamefont {D.~J.}\
      \bibnamefont {Wineland}}, \ and\ \bibinfo {author} {\bibfnamefont {D.~J.}\
      \bibnamefont {Heinzen}},\ }\bibfield {title} {\emph {\enquote {\bibinfo
        {title} {Optimal frequency measurements with maximally correlated
          states},}\ }}\href {\doibase 10.1103/PhysRevA.54.R4649}
  {\bibfield {journal} {\bibinfo {journal} {Phys. Rev. A}\ }\textbf {\bibinfo
      {volume} {54}},\ \bibinfo {pages} {R4649} (\bibinfo {year}
    {1996})}\BibitemShut {NoStop}%
\bibitem [{\citenamefont {Huelga}\ \emph {et~al.}(1997)\citenamefont {Huelga},
    \citenamefont {Macchiavello}, \citenamefont {Pellizzari}, \citenamefont
    {Ekert}, \citenamefont {Plenio},\ and\ \citenamefont {Cirac}}]{Cirac1997}%
  \BibitemOpen \bibfield {author} {\bibinfo {author} {\bibfnamefont {S.~F.}\
      \bibnamefont {Huelga}}, \bibinfo {author} {\bibfnamefont {C.}~\bibnamefont
      {Macchiavello}}, \bibinfo {author} {\bibfnamefont {T.}~\bibnamefont
      {Pellizzari}}, \bibinfo {author} {\bibfnamefont {A.~K.}\ \bibnamefont
      {Ekert}}, \bibinfo {author} {\bibfnamefont {M.~B.}\ \bibnamefont
      {Plenio}}, \ and\ \bibinfo {author} {\bibfnamefont {J.~I.}\ \bibnamefont
      {Cirac}},\ }\bibfield {title} {\emph {\enquote {\bibinfo {title}
        {Improvement of frequency standards with quantum entanglement},}\
    }}\href {\doibase 10.1103/PhysRevLett.79.3865}
  {\bibfield {journal} {\bibinfo {journal} {Phys. Rev. Lett.}\ }\textbf
    {\bibinfo {volume} {79}},\ \bibinfo {pages} {3865} (\bibinfo {year}
    {1997})}\BibitemShut {NoStop}%
\bibitem [{\citenamefont {Cohen}\ \emph {et~al.}(2020)\citenamefont {Cohen},
    \citenamefont {Gefen}, \citenamefont {Ortiz},\ and\ \citenamefont
    {Retzker}}]{Cohen2020}%
  \BibitemOpen \bibfield {author} {\bibinfo {author} {\bibfnamefont
      {D.}~\bibnamefont {Cohen}}, \bibinfo {author} {\bibfnamefont
      {T.}~\bibnamefont {Gefen}}, \bibinfo {author} {\bibfnamefont
      {L.}~\bibnamefont {Ortiz}}, \ and\ \bibinfo {author} {\bibfnamefont
      {A.}~\bibnamefont {Retzker}},\ }\bibfield {title} {\emph {\enquote
      {\bibinfo {title} {Achieving the ultimate precision limit with a weakly
          interacting quantum probe},}\ }}\href
  {\doibase 10.1038/s41534-020-00313-x} {\bibfield {journal} {\bibinfo {journal}
      {npj Quantum Information}\ }\textbf {\bibinfo {volume} {6}},\ \bibinfo
    {pages} {1} (\bibinfo {year} {2020})}\BibitemShut {NoStop}%
\bibitem [{\citenamefont {Toscano}\ \emph {et~al.}(2017)\citenamefont
  {Toscano}, \citenamefont {Bastos},\ and\ \citenamefont
  {de~Matos~Filho}}]{Toscano2017}%
  \BibitemOpen
  \bibfield  {author} {\bibinfo {author} {\bibfnamefont {F.}~\bibnamefont
  {Toscano}}, \bibinfo {author} {\bibfnamefont {W.~P.}\ \bibnamefont {Bastos}},
  \ and\ \bibinfo {author} {\bibfnamefont {R.~L.}\ \bibnamefont
  {de~Matos~Filho}},\ }\bibfield  {title} {\emph {\enquote {\bibinfo {title}
  {Attainability of the quantum information bound in pure-state models},}\
  }}\href {\doibase 10.1103/PhysRevA.95.042125} {\bibfield  {journal} {\bibinfo
   {journal} {Phys. Rev. A}\ }\textbf {\bibinfo {volume} {95}},\ \bibinfo
  {pages} {042125} (\bibinfo {year} {2017})}\BibitemShut {NoStop}%
\bibitem [{\citenamefont {Bonato}\ \emph {et~al.}(2016)\citenamefont {Bonato},
    \citenamefont {Blok}, \citenamefont {Dinani}, \citenamefont {Berry},
    \citenamefont {Markham}, \citenamefont {Twitchen},\ and\ \citenamefont
    {Hanson}}]{bonato2016optimized}%
  \BibitemOpen \bibfield {author} {\bibinfo {author} {\bibfnamefont
      {C.}~\bibnamefont {Bonato}}, \bibinfo {author} {\bibfnamefont {M.~S.}\
      \bibnamefont {Blok}}, \bibinfo {author} {\bibfnamefont {H.~T.}\
      \bibnamefont {Dinani}}, \bibinfo {author} {\bibfnamefont {D.~W.}\
      \bibnamefont {Berry}}, \bibinfo {author} {\bibfnamefont {M.~L.}\
      \bibnamefont {Markham}}, \bibinfo {author} {\bibfnamefont {D.~J.}\
      \bibnamefont {Twitchen}}, \ and\ \bibinfo {author} {\bibfnamefont
      {R.}~\bibnamefont {Hanson}},\ }\bibfield {title} {\emph {\enquote
      {\bibinfo {title} {Optimized quantum sensing with a single electron spin
          using real-time adaptive measurements},}\ }}\href
  {\doibase 10.1038/nnano.2015.261} {\bibfield {journal} {\bibinfo {journal}
      {Nature nanotechnology}\ }\textbf {\bibinfo {volume} {11}},\ \bibinfo
    {pages} {247} (\bibinfo {year} {2016})}\BibitemShut {NoStop}%
\bibitem [{\citenamefont {Keener}(2010)}]{Robert}%
  \BibitemOpen
  \bibfield  {author} {\bibinfo {author} {\bibfnamefont {R.~W.}\ \bibnamefont
  {Keener}},\ }\href {\doibase 10.1007/978-0-387-93839-4} {\emph {\bibinfo
  {title} {Theoretical Statistics: Topics for a Core Course}}},\ Springer Texts
  in Statistics\ (\bibinfo  {publisher} {Springer New York},\ \bibinfo {year}
  {2010})\BibitemShut {NoStop}%
\bibitem [{\citenamefont {Lehmann}\ and\ \citenamefont
  {Casella}(2006)}]{Lehmann1998}%
  \BibitemOpen
  \bibfield  {author} {\bibinfo {author} {\bibfnamefont {E.}~\bibnamefont
  {Lehmann}}\ and\ \bibinfo {author} {\bibfnamefont {G.}~\bibnamefont
  {Casella}},\ }\href {https://books.google.com/books?id=4f24CgAAQBAJ} {\emph
  {\bibinfo {title} {Theory of Point Estimation}}},\ Springer Texts in
  Statistics\ (\bibinfo  {publisher} {Springer New York},\ \bibinfo {year}
  {2006})\BibitemShut {NoStop}%
\bibitem [{\citenamefont {Greenberger}\ \emph {et~al.}(1989)\citenamefont
  {Greenberger}, \citenamefont {Horne},\ and\ \citenamefont
  {Zeilinger}}]{greenberger1989going}%
  \BibitemOpen
  \bibfield  {author} {\bibinfo {author} {\bibfnamefont {D.~M.}\ \bibnamefont
  {Greenberger}}, \bibinfo {author} {\bibfnamefont {M.~A.}\ \bibnamefont
  {Horne}}, \ and\ \bibinfo {author} {\bibfnamefont {A.}~\bibnamefont
  {Zeilinger}},\ }in\ \href {\doibase 10.1007/978-94-017-0849-4_10} {\emph
  {\bibinfo {booktitle} {Bell's Theorem, Quantum Theory and Conceptions of the
  Universe}}},\ \bibinfo {editor} {edited by\ \bibinfo {editor} {\bibfnamefont
  {M.}~\bibnamefont {Kafatos}}}\ (\bibinfo  {publisher} {Springer
  Netherlands},\ \bibinfo {address} {Dordrecht},\ \bibinfo {year} {1989})\ pp.\
  \bibinfo {pages} {69--72}\BibitemShut {NoStop}%
\bibitem [{\citenamefont {Escher}\ \emph {et~al.}(2011)\citenamefont {Escher},
    \citenamefont {de~Matos~Filho},\ and\ \citenamefont
    {Davidovich}}]{escher2011quantum}%
  \BibitemOpen \bibfield {author} {\bibinfo {author} {\bibfnamefont {B.~M.}\
      \bibnamefont {Escher}}, \bibinfo {author} {\bibfnamefont {R.~L.}\
      \bibnamefont {de~Matos~Filho}}, \ and\ \bibinfo {author} {\bibfnamefont
      {L.}~\bibnamefont {Davidovich}},\ }\bibfield {title} {\emph {\enquote
      {\bibinfo {title} {Quantum metrology for noisy systems},}\ }}\href
  {\doibase 10.1007/s13538-011-0037-y} {\bibfield {journal} {\bibinfo {journal}
      {Brazilian Journal of Physics}\ }\textbf {\bibinfo {volume} {41}},\
    \bibinfo {pages} {229} (\bibinfo {year} {2011})}\BibitemShut {NoStop}%
\bibitem [{\citenamefont {Fujiwara}(2006)}]{Fujiwara2011}%
  \BibitemOpen
  \bibfield  {author} {\bibinfo {author} {\bibfnamefont {A.}~\bibnamefont
  {Fujiwara}},\ }\bibfield  {title} {\emph {\enquote {\bibinfo {title} {Strong
  consistency and asymptotic efficiency for adaptive quantum estimation
  problems},}\ }}\href {\doibase 10.1088/0305-4470/39/40/014} {\bibfield
  {journal} {\bibinfo  {journal} {Journal of Physics A: Mathematical and
  General}\ }\textbf {\bibinfo {volume} {39}},\ \bibinfo {pages} {12489}
  (\bibinfo {year} {2006})}\BibitemShut {NoStop}%
\bibitem [{\citenamefont {Rodr{\'{i}}guez-Garc{\'{i}}a}\ \emph
  {et~al.}(2021)\citenamefont {Rodr{\'{i}}guez-Garc{\'{i}}a}, \citenamefont
  {Castillo},\ and\ \citenamefont
  {Barberis-Blostein}}]{RodriguezGarcia2021efficientqubitphase}%
  \BibitemOpen
  \bibfield  {author} {\bibinfo {author} {\bibfnamefont {M.~A.}\ \bibnamefont
  {Rodr{\'{i}}guez-Garc{\'{i}}a}}, \bibinfo {author} {\bibfnamefont {I.~P.}\
  \bibnamefont {Castillo}}, \ and\ \bibinfo {author} {\bibfnamefont
  {P.}~\bibnamefont {Barberis-Blostein}},\ }\bibfield  {title} {\emph {\enquote
  {\bibinfo {title} {Efficient qubit phase estimation using adaptive
  measurements},}\ }}\href {\doibase 10.22331/q-2021-06-04-467} {\bibfield
  {journal} {\bibinfo  {journal} {{Quantum}}\ }\textbf {\bibinfo {volume}
  {5}},\ \bibinfo {pages} {467} (\bibinfo {year} {2021})}\BibitemShut {NoStop}%
\bibitem [{\citenamefont {Pezzé}\ and\ \citenamefont
  {Smerzi}(2007)}]{pezzesmerzi2007}%
  \BibitemOpen
  \bibfield  {author} {\bibinfo {author} {\bibfnamefont {L.}~\bibnamefont
  {Pezzé}}\ and\ \bibinfo {author} {\bibfnamefont {A.}~\bibnamefont
  {Smerzi}},\ }\bibfield  {title} {\emph {\enquote {\bibinfo {title} {Sub
  shot-noise interferometric phase sensitivity with beryllium ions schrödinger
  cat states},}\ }}\href {\doibase 10.1209/0295-5075/78/30004} {\bibfield
  {journal} {\bibinfo  {journal} {Europhysics Letters}\ }\textbf {\bibinfo
  {volume} {78}},\ \bibinfo {pages} {30004} (\bibinfo {year}
  {2007})}\BibitemShut {NoStop}%
\bibitem [{\citenamefont {Oh}\ and\ \citenamefont {Son}(2014)}]{Oh_2014}%
  \BibitemOpen \bibfield {author} {\bibinfo {author} {\bibfnamefont
      {C.}~\bibnamefont {Oh}}\ and\ \bibinfo {author} {\bibfnamefont
      {W.}~\bibnamefont {Son}},\ }\bibfield {title} {\emph {\enquote {\bibinfo
        {title} {Sub shot-noise frequency estimation with bounded a priori
          knowledge},}\ }}\href {\doibase 10.1088/1751-8113/48/4/045304}
  {\bibfield {journal} {\bibinfo {journal} {Journal of Physics A: Mathematical
        and Theoretical}\ }\textbf {\bibinfo {volume} {48}},\ \bibinfo {pages}
    {045304} (\bibinfo {year} {2014})}\BibitemShut {NoStop}%
\bibitem [{\citenamefont {Berry}\ \emph {et~al.}(2009)\citenamefont {Berry},
  \citenamefont {Higgins}, \citenamefont {Bartlett}, \citenamefont {Mitchell},
  \citenamefont {Pryde},\ and\ \citenamefont {Wiseman}}]{Berry2009}%
  \BibitemOpen
  \bibfield  {author} {\bibinfo {author} {\bibfnamefont {D.~W.}\ \bibnamefont
  {Berry}}, \bibinfo {author} {\bibfnamefont {B.~L.}\ \bibnamefont {Higgins}},
  \bibinfo {author} {\bibfnamefont {S.~D.}\ \bibnamefont {Bartlett}}, \bibinfo
  {author} {\bibfnamefont {M.~W.}\ \bibnamefont {Mitchell}}, \bibinfo {author}
  {\bibfnamefont {G.~J.}\ \bibnamefont {Pryde}}, \ and\ \bibinfo {author}
  {\bibfnamefont {H.~M.}\ \bibnamefont {Wiseman}},\ }\bibfield  {title} {\emph
  {\enquote {\bibinfo {title} {How to perform the most accurate possible phase
  measurements},}\ }}\href {\doibase 10.1103/PhysRevA.80.052114} {\bibfield
  {journal} {\bibinfo  {journal} {Phys. Rev. A}\ }\textbf {\bibinfo {volume}
  {80}},\ \bibinfo {pages} {052114} (\bibinfo {year} {2009})}\BibitemShut
  {NoStop}%
\bibitem [{\citenamefont {Paninski}(2005)}]{paninski2005asymptotic}%
  \BibitemOpen
  \bibfield  {author} {\bibinfo {author} {\bibfnamefont {L.}~\bibnamefont
  {Paninski}},\ }\bibfield  {title} {\emph {\enquote {\bibinfo {title}
  {{Asymptotic Theory of Information-Theoretic Experimental Design}},}\ }}\href
  {\doibase 10.1162/0899766053723032} {\bibfield  {journal} {\bibinfo
  {journal} {Neural Computation}\ }\textbf {\bibinfo {volume} {17}},\ \bibinfo
  {pages} {1480} (\bibinfo {year} {2005})}\BibitemShut {NoStop}%
\bibitem [{\citenamefont {Boixo}\ and\ \citenamefont {Somma}(2008)}]{Boxio}%
  \BibitemOpen \bibfield {author} {\bibinfo {author} {\bibfnamefont
      {S.}~\bibnamefont {Boixo}}\ and\ \bibinfo {author} {\bibfnamefont {R.~D.}\
      \bibnamefont {Somma}},\ }\bibfield {title} {\emph {\enquote {\bibinfo
        {title} {Parameter estimation with mixed-state quantum computation},}\
    }}\href {\doibase 10.1103/PhysRevA.77.052320} {\bibfield {journal} {\bibinfo
      {journal} {Phys. Rev. A}\ }\textbf {\bibinfo {volume} {77}},\
    \bibinfo {pages} {052320} (\bibinfo {year} {2008})}\BibitemShut {NoStop}%
\bibitem [{\citenamefont {Huang}\ \emph {et~al.}(2017)\citenamefont {Huang},
    \citenamefont {Motes}, \citenamefont {Anisimov}, \citenamefont {Dowling},\
    and\ \citenamefont {Berry}}]{Huang2017}%
  \BibitemOpen \bibfield {author} {\bibinfo {author} {\bibfnamefont
      {Z.}~\bibnamefont {Huang}}, \bibinfo {author} {\bibfnamefont {K.~R.}\
      \bibnamefont {Motes}}, \bibinfo {author} {\bibfnamefont {P.~M.}\
      \bibnamefont {Anisimov}}, \bibinfo {author} {\bibfnamefont {J.~P.}\
      \bibnamefont {Dowling}}, \ and\ \bibinfo {author} {\bibfnamefont {D.~W.}\
      \bibnamefont {Berry}},\ }\bibfield {title} {\emph {\enquote {\bibinfo
        {title} {Adaptive phase estimation with two-mode squeezed vacuum and
          parity measurement},}\ }}\href {\doibase 10.1103/PhysRevA.95.053837}
  {\bibfield {journal} {\bibinfo {journal} {Phys. Rev. A}\ }\textbf {\bibinfo
      {volume} {95}},\ \bibinfo {pages} {053837} (\bibinfo {year}
    {2017})}\BibitemShut {NoStop}%
\bibitem [{\citenamefont {Berry}\ and\ \citenamefont
  {Wiseman}(2001)}]{Berry2001}%
  \BibitemOpen
  \bibfield  {author} {\bibinfo {author} {\bibfnamefont {D.}~\bibnamefont
  {Berry}}\ and\ \bibinfo {author} {\bibfnamefont {H.}~\bibnamefont
  {Wiseman}},\ }in\ \href {\doibase 10.1109/QELS.2001.961853} {\emph {\bibinfo
  {booktitle} {Technical Digest. Summaries of papers presented at the Quantum
  Electronics and Laser Science Conference. Postconference Technical Digest
  (IEEE Cat. No.01CH37172)}}}\ (\bibinfo {year} {2001})\ pp.\ \bibinfo {pages}
  {60--61}\BibitemShut {NoStop}%
\bibitem [{\citenamefont {Rodr{\'i}guez-Garc{\'i}a}\ \emph
  {et~al.}(2022)\citenamefont {Rodr{\'i}guez-Garc{\'i}a}, \citenamefont
  {DiMario}, \citenamefont {Barberis-Blostein},\ and\ \citenamefont
  {Becerra}}]{rodriguez2022determination}%
  \BibitemOpen
  \bibfield  {author} {\bibinfo {author} {\bibfnamefont {M.~A.}\ \bibnamefont
  {Rodr{\'i}guez-Garc{\'i}a}}, \bibinfo {author} {\bibfnamefont {M.~T.}\
  \bibnamefont {DiMario}}, \bibinfo {author} {\bibfnamefont {P.}~\bibnamefont
  {Barberis-Blostein}}, \ and\ \bibinfo {author} {\bibfnamefont {F.~E.}\
  \bibnamefont {Becerra}},\ }\bibfield  {title} {\emph {\enquote {\bibinfo
  {title} {Determination of the asymptotic limits of adaptive photon counting
  measurements for coherent-state optical phase estimation},}\ }}\href
  {\doibase 10.1038/s41534-022-00601-8} {\bibfield  {journal} {\bibinfo
  {journal} {npj Quantum Information}\ }\textbf {\bibinfo {volume} {8}},\
  \bibinfo {pages} {94} (\bibinfo {year} {2022})}\BibitemShut {NoStop}%
\bibitem [{\citenamefont {Holevo}(2011)}]{Holevo2011}%
  \BibitemOpen
  \bibfield  {author} {\bibinfo {author} {\bibfnamefont {A.}~\bibnamefont
  {Holevo}},\ }\href {\doibase 10.1007/978-88-7642-378-9} {\emph {\bibinfo
  {title} {Probabilistic and Statistical Aspects of Quantum Theory}}}\
  (\bibinfo  {publisher} {Edizioni della Normale},\ \bibinfo {year}
  {2011})\BibitemShut {NoStop}%
\bibitem [{\citenamefont {Helstrom}(1969)}]{helstrom1969quantum}%
  \BibitemOpen
  \bibfield  {author} {\bibinfo {author} {\bibfnamefont {C.~W.}\ \bibnamefont
  {Helstrom}},\ }\bibfield  {title} {\emph {\enquote {\bibinfo {title} {Quantum
  detection and estimation theory},}\ }}\href {\doibase 10.1007/bf01007479}
  {\bibfield  {journal} {\bibinfo  {journal} {Journal of Statistical Physics}\
  }\textbf {\bibinfo {volume} {1}},\ \bibinfo {pages} {231–252} (\bibinfo
  {year} {1969})}\BibitemShut {NoStop}%
\bibitem [{\citenamefont {Beneduci}(2011)}]{beneduci2011relationships}%
  \BibitemOpen
  \bibfield  {author} {\bibinfo {author} {\bibfnamefont {R.}~\bibnamefont
  {Beneduci}},\ }\bibfield  {title} {\emph {\enquote {\bibinfo {title} {On the
  relationships between the moments of a povm and the generator of the von
  neumann algebra it generates},}\ }}\href {\doibase 10.1007/s10773-011-0907-7}
  {\bibfield  {journal} {\bibinfo  {journal} {International Journal of
  Theoretical Physics}\ }\textbf {\bibinfo {volume} {50}},\ \bibinfo {pages}
  {3724–3736} (\bibinfo {year} {2011})}\BibitemShut {NoStop}%
\bibitem [{\citenamefont {DeGroot}\ and\ \citenamefont
  {Schervish}(2013)}]{degroot}%
  \BibitemOpen
  \bibfield  {author} {\bibinfo {author} {\bibfnamefont {M.}~\bibnamefont
  {DeGroot}}\ and\ \bibinfo {author} {\bibfnamefont {M.}~\bibnamefont
  {Schervish}},\ }\href {https://books.google.com/books?id=hIPkngEACAAJ} {\emph
  {\bibinfo {title} {Probability and Statistics}}},\ Pearson custom library\
  (\bibinfo  {publisher} {Pearson Education},\ \bibinfo {year}
  {2013})\BibitemShut {NoStop}%
\bibitem [{\citenamefont {Casella}\ and\ \citenamefont
  {Berger}(2024)}]{CaseBerg:01}%
  \BibitemOpen
  \bibfield  {author} {\bibinfo {author} {\bibfnamefont {G.}~\bibnamefont
  {Casella}}\ and\ \bibinfo {author} {\bibfnamefont {R.}~\bibnamefont
  {Berger}},\ }\href {\doibase 10.1201/9781003456285} {\emph {\bibinfo {title}
  {Statistical Inference}}}\ (\bibinfo  {publisher} {Chapman and Hall/CRC},\
  \bibinfo {year} {2024})\BibitemShut {NoStop}%
\bibitem [{\citenamefont {Chapeau-Blondeau}(2016)}]{Chapeau2016}%
  \BibitemOpen \bibfield {author} {\bibinfo {author} {\bibfnamefont
      {F.}~\bibnamefont {Chapeau-Blondeau}},\ }\bibfield {title} {\emph
    {\enquote {\bibinfo {title} {{Optimizing qubit phase estimation}},}\ }}\href
  {\doibase 10.1103/PhysRevA.94.022334} {\bibfield {journal} {\bibinfo {journal}
      {Physical Review A}\ }\textbf {\bibinfo {volume} {94}},\ \bibinfo {pages}
    {1} (\bibinfo {year} {2016})}\BibitemShut {NoStop}%
\bibitem [{\citenamefont {Holevo}(1979)}]{Holevo1979}%
  \BibitemOpen \bibfield {author} {\bibinfo {author} {\bibfnamefont
      {A.}~\bibnamefont {Holevo}},\ }\bibfield {title} {\emph {\enquote
      {\bibinfo {title} {Covariant measurements and uncertainty relations},}\
    }}\href {\doibase 10.1016/0034-4877(79)90072-7}
  {\bibfield {journal} {\bibinfo {journal} {Reports on Mathematical Physics}\
    }\textbf {\bibinfo {volume} {16}},\ \bibinfo {pages} {385–400} (\bibinfo
    {year} {1979})}\BibitemShut {NoStop}%
\bibitem [{\citenamefont {Martin}\ \emph {et~al.}(2020)\citenamefont {Martin},
  \citenamefont {Livingston}, \citenamefont {Hacohen-Gourgy}, \citenamefont
  {Wiseman},\ and\ \citenamefont {Siddiqi}}]{Martin2019}%
  \BibitemOpen
  \bibfield  {author} {\bibinfo {author} {\bibfnamefont {L.~S.}\ \bibnamefont
  {Martin}}, \bibinfo {author} {\bibfnamefont {W.~P.}\ \bibnamefont
  {Livingston}}, \bibinfo {author} {\bibfnamefont {S.}~\bibnamefont
  {Hacohen-Gourgy}}, \bibinfo {author} {\bibfnamefont {H.~M.}\ \bibnamefont
  {Wiseman}}, \ and\ \bibinfo {author} {\bibfnamefont {I.}~\bibnamefont
  {Siddiqi}},\ }\bibfield  {title} {\emph {\enquote {\bibinfo {title}
  {Implementation of a canonical phase measurement with quantum feedback},}\
  }}\href {\doibase 10.1038/s41567-020-0939-0} {\bibfield  {journal} {\bibinfo
  {journal} {Nature Physics}\ }\textbf {\bibinfo {volume} {16}},\ \bibinfo
  {pages} {1046–1049} (\bibinfo {year} {2020})}\BibitemShut {NoStop}%
\bibitem [{\citenamefont {Okamoto}\ \emph {et~al.}(2017)\citenamefont {Okamoto},
    \citenamefont {Oyama}, \citenamefont {Yamagata}, \citenamefont {Fujiwara},\
    and\ \citenamefont {Takeuchi}}]{Fuji2}%
  \BibitemOpen \bibfield {author} {\bibinfo {author} {\bibfnamefont
      {R.}~\bibnamefont {Okamoto}}, \bibinfo {author} {\bibfnamefont
      {S.}~\bibnamefont {Oyama}}, \bibinfo {author} {\bibfnamefont
      {K.}~\bibnamefont {Yamagata}}, \bibinfo {author} {\bibfnamefont
      {A.}~\bibnamefont {Fujiwara}}, \ and\ \bibinfo {author} {\bibfnamefont
      {S.}~\bibnamefont {Takeuchi}},\ }\bibfield {title} {\emph {\enquote
      {\bibinfo {title} {Experimental demonstration of adaptive quantum state
          estimation for single photonic qubits},}\ }}\href
  {\doibase 10.1103/PhysRevA.96.022124} {\bibfield {journal} {\bibinfo {journal}
      {Phys. Rev. A}\ }\textbf {\bibinfo {volume} {96}},\ \bibinfo {pages}
    {022124} (\bibinfo {year} {2017})}\BibitemShut {NoStop}%
\bibitem [{\citenamefont {Shaw}\ \emph {et~al.}(2024)\citenamefont {Shaw},
    \citenamefont {Finkelstein}, \citenamefont {Tsai}, \citenamefont {Scholl},
    \citenamefont {Yoon}, \citenamefont {Choi},\ and\ \citenamefont
    {Endres}}]{Shaw2024}%
  \BibitemOpen \bibfield {author} {\bibinfo {author} {\bibfnamefont {A.~L.}\
      \bibnamefont {Shaw}}, \bibinfo {author} {\bibfnamefont {R.}~\bibnamefont
      {Finkelstein}}, \bibinfo {author} {\bibfnamefont {R.~B.-S.}\ \bibnamefont
      {Tsai}}, \bibinfo {author} {\bibfnamefont {P.}~\bibnamefont {Scholl}},
    \bibinfo {author} {\bibfnamefont {T.~H.}\ \bibnamefont {Yoon}}, \bibinfo
    {author} {\bibfnamefont {J.}~\bibnamefont {Choi}}, \ and\ \bibinfo {author}
    {\bibfnamefont {M.}~\bibnamefont {Endres}},\ }\bibfield {title} {\emph
    {\enquote {\bibinfo {title} {Multi-ensemble metrology by programming local
          rotations with atom movements},}\ }}\href
  {\doibase 10.1038/s41567-023-02323-w} {\bibfield {journal} {\bibinfo {journal}
      {Nature Physics}\ }\textbf {\bibinfo {volume} {20}},\ \bibinfo {pages}
    {195–201} (\bibinfo {year} {2024})}\BibitemShut {NoStop}%
\end{thebibliography}

%

\end{document}